\documentclass[aps,prb,twocolumn,secnumarabic,amsmath,amssymb,superscriptaddress]{revtex4-1}
\usepackage{dcolumn}
\usepackage{bm}

\usepackage{color} 
\usepackage{ulem}

\usepackage{amsmath}
\usepackage{graphicx}
\usepackage{float}
\usepackage{subfig}

\usepackage{color}
\usepackage[colorlinks,bookmarks=false,citecolor=blue,linkcolor=red,urlcolor=blue]{hyperref}

\definecolor{darkred}{rgb}{0.7,0.0,0.0}

\definecolor{darkblue}{rgb}{0,0.02,0.45}

\definecolor{darkgreen}{rgb}{0.02,0.45,0.0}

\definecolor{violet}{rgb}{0.8,0.2,0.6}

\providecommand{\U}[1]{\protect\rule{.1in}{.1in}}

\begin{document}

\title{Properties and application of the SO(3) Majorana representation of spin: equivalence with the Jordan-Wigner transformation and exact $Z_{2}$ gauge theories for spin models}

\author{Jianlong Fu}
\affiliation{School of Physics and Astronomy, University of Minnesota, Minneapolis, Minnesota 55455, USA}

\begin{abstract}
We explore the properties of the SO(3) Majorana representation of spin. Based on its non-local nature, it is shown that there is an equivalence between the SO(3) Majorana representation and the Jordan-Wigner transformation in one and two dimensions. From the relation between the SO(3) Majorana representation and one-dimensional Jordan-Wigner transformation, we show that application of the SO(3) Majorana representation usually results in $Z_{2}$ gauge structure. Based on lattice Chern-Simons gauge theory, it is shown that the anti-commuting link variables in the SO(3) Majorana representation make it equivalent to an operator form of compact $\text{U(1)}_{1}$ Chern-Simons Jordan-Wigner transformation in 2d. As examples of its application, we discuss two spin models, namely the quantum XY model on honeycomb lattice and the $90^{\circ}$ compass model on square lattice. It is shown that under the SO(3) Majorana representation both spin models can be exactly mapped into $Z_{2}$ gauge theory of spinons, with the standard form of $Z_{2}$ Gauss law constraint.    
\end{abstract}

\maketitle
\section{Introduction}

The study of quantum spin systems always involves representation of spin in terms of bosonic or fermionic quasi-particles, which are often called spinons \cite{auerbachbook,Savary2016}. Quantum dynamics of such spinons offers insights in the properties of the spin system. In particular, we are interested in the ground states of spin systems. If the original spin ground states breaks spin rotational symmetry, i.e. the state is ordered, then the spectrum of the spin systems can be described by spin-wave excitations \cite{auerbachbook}, carrying spin-1 if the original spin system is spin-$\frac{1}{2}$. These excitations are usually seen as confined pairs of the spinons. Such behaviour is usually found in ordinary magnetic materials. There are other types of spin ground states which do not break any symmetry, such ground states are called quantum spin liquid (QSL) states \cite{Savary2016,Balents2010,Zhou2017,wen1991,anderson1973,fazekas1974}. QSL states show non-trivial quantum entanglement \cite{Savary2016}, and possess topological degeneracy \cite{wen1991}; their excitations are usually described by weakly interacting spinons. While the theoretical description of ordered spin states is rather simple, the study of quantum spin liquid states is much more involved and results in the introduction of many types of spin representations \cite{Savary2016,fu2018,batista2004}.

Some of the representations start from a local mapping between spin operators and bilinear form of quasi-particle operators. Here we give a few examples. The most commonly used one of this kind is the {\it Abrikosov fermion} representation \cite{marston1989,wen1991,Lee2006,Savary2016}, it defines two complex fermions to represent spin operator. Due to the fact that the local Hilbert space of two fermions is two times larger than the spin Hilbert space, a single-occupation constraint must be added to ensure that the spin space is faithfully represented \cite{Lee2006}. Besides the Abrikosov fermion representation, various types of Majorana fermion representation have also been introduced. Some of these use four Majorana fermions to represent each spin, including the {\it Kitaev representaion} \cite{Kitaev2006} and the {\it SO(4) chiral Majorana representation} \cite{fu2018,Chen2012}. The Hilbert space of four Majorana fermions sharing the same spatial position can be defined locally since it is possible to pair them up into two pairs and define two complex fermions. Actually the local Hilbert space of the four Majorana fermions in the Kitaev and the SO(4) chiral Majorana representation is the same as the one of the Abrikosov fermion representation \cite{fu2018}. Besides, special representations have been defined and applied to lattice spin models with exotic geometry, such as the quantum spin ice model \cite{savary2012,lee2012,fu2017}.

Other types of representation define a non-local mapping between spin operators and fermionic (or possibly bosonic) operators. The {\it Jordan-Wigner transformation}  \cite{jordan1928,lieb1961,fradkinbook} defines a one-dimensional (1d) mapping between spin operators in a spin chain to a fermion attached to a half-infinite string operator which creats a quantum kink. Using this transformation the one-dimensional quantum XY model is mapped into a free fermion hopping model, which is exactly solvable. The generalization of the Jordan-Wigner transformation to higher dimensions is also available \cite{tsvelikbook}. In particular, the Jordan-Wigner transformation in two-dimensions (2d) involves Chern-Simons (CS) gauge theory \cite{fradkin1989,lopez1994,fradkinbook}, and thus it is often called {\it Chern-Simons Jordan-Wigner transformation}. Defining a mapping between spin operators and a fermion coupled to a string-operator of Chern-Simons gauge field \cite{kumar2014}, the 2d Jordan-Wigner transformation maps the quantum XY model in 2d into a model of complex fermion coupled to a Chern-Simons gauge theory (Some details of the 2d Jordan-Wigner transformation are given in Appendix \ref{appendix2djordanwigner}). Besides the Jordan-Wigner (JW) transformation, there is another non-local representation of spin, which is the {\it SO(3) Majorana representation} \cite{Berezin1975,Berezin1977,Tsvelik1992,tsvelikbook,Shnirman2003,Mao2003,fu2018,Herfurth2013,Shastry1997,Biswas2011}. In the SO(3) Majorana representation, each spin operator is represented by three Majorana fermions. Because the number of Majorana fermions defined on each site is odd, it is not possible to establish the Majorana Hilbert space locally. Instead, we have to pair up sites and the Majorana fermions on them to define the Majorana Hilbert space \cite{Biswas2011,fu2018}. This pairing results in the non-local nature of the SO(3) Majorana representation of spin. 

In this work, we focus on the properties and the application of the SO(3) Majorana representation of spin. Knowing its non-local nature, our first question to ask is whether there is any relation between the SO(3) Majorana representation and the Jordan-Wigner transformation in 1d and 2d. If there is such relation, what will it tell us about the properties of the SO(3) Majorana representation? To answer these questions, in Sec. \ref{sec1djordanwigner} we discuss the relation between the SO(3) Majorana representation and the 1d Jordan-Wigner transformation in the one-dimensional spin chain. We argue that under some specific conditions introduced to fix the Majorana Hilbert space, the SO(3) Majorana representation of spin can be mapped into the 1d Jordan-Wigner transformation. In the discussion, we also show that there should always be some $Z_{2}$ redundancy if we only impose $\frac{N}{2}$ ($N$ is the total number of spins in the system) conditions to fix the Majorana Hilbert space. In Sec. \ref{sec2djordanwigner}, we analyze the relationship between the SO(3) Majorana representation and the Chern-Simons JW transformation in the two-dimensional XXZ Heisenberg model. With the proper definition of the lattice Chern-Simons gauge theory \cite{sun2015}, we show that the SO(3) Majorana representation can be seen as an operator form of the Chern-Simons JW transformation due to the existence of anti-commuting link variables in both representations. Furthermore, we argue that the gauge field in the lattice Chern-Simons Jordan-Wigner transformation is compact which leads to the quantization of the CS gauge connection. Such quantization further confirms the correspondence between the SO(3) Majorana representation and Chern-Simons Jordan-Wigner transformation. The correspondence we find in the specific models can be directly generalized to other spin models. It means that, except for some technical details which will be explained later, the application of the SO(3) Majorana representation and the Jordan-Wigner transformation in one and two dimensions are equivalent to each other physically in any spin models. 

In light of these, our second goal of this work is to explore the application of the SO(3) Majorana representation in various spin models. Previous studies of spin models using various spin representations usually results in some lattice gauge theory \cite{Kogut1979,Prosko2017,senthil2000}. In the Abrikosov fermion representation, starting from mean field treatment of the quartic interacting terms of the spinons, the gauge structure (which is SU(2)) emerges after neglecting the fluctuation of the modulus of the Hubbard-Stratonovich field but keeping the fluctuation of its phases \cite{Lee2006,wen1991,Savary2016}. Application of non-local 2d Jordan-Wigner transformation results in U(1) Chern-Simons gauge theories \cite{fradkin1989,lopez1994,kumar2014}. Mean field studies of various spin models \cite{Herfurth2013,Biswas2011} and exact solution of Kitaev model \cite{fu2018} using the SO(3) Majorana representation result in $Z_{2}$ gauge theories. Thus, it is conjectured that the application of the SO(3) Majorana representation can result in $Z_{2}$ lattice gauge theories because of the $Z_{2}$ redundancy mentioned above.  

Inspired by the method used in the solution of the Kitaev model in Ref. \onlinecite{fu2018}, we consider two spin models in this work, namely the quantum XY model on honeycomb lattice and the $90^{\circ}$ compass model on square lattice using the SO(3) Majorana representation. Our results confirm the appearing of $Z_{2}$ lattice gauge theory in both models and illustrate how to obtain the lattice $Z_{2}$ gauge theory from various exact transformations. We also show that the conditions we apply to fix the Majorana Hilbert space can be mapped into standard form of Gauss law in $Z_{2}$ gauge theory. To this end, our study is different from previous studies \cite{Herfurth2013,Biswas2011} in that no approximation is introduced in obtaining the $Z_{2}$ gauge theories. Unfortunately the resulting $Z_{2}$ gauge theory does not take the standard form \cite{Prosko2017} and contain some non-trivial features. Some approximations are needed to treat these $Z_{2}$ gauge theories. Although neither of these models is exactly solvable, the way we obtain the $Z_{2}$ lattice gauge theory may open a window to a new perspective on the study of the spin systems. Some discussions about these considerations are given in later sections.

The rest of the paper is organized as follows. In Sec. \ref{secso3}, we introduce the SO(3) Majorana representation of spin and review its basic properties. In Sec. \ref{sec1djordanwigner} and Sec. \ref{sec2djordanwigner}, we discuss the relation between the SO(3) Majorana representation and the Jordan-Wigner transformation in 1d and 2d respectively. In Sec. \ref{sectwospinmodels}, we discuss the application of SO(3) Majorana representation in two spin models, namely the quantum XY model on honeycomb lattice (in Sec. \ref{secxyhoneycomb}) and the $90^{\circ}$ compass model on square lattice (in Sec. \ref{seccompass}). A discussion on the results and further application is given in Sec. \ref{secdiscussiononapplication}. The paper concludes in Sec. \ref{secconclusion} with some open questions and direction for future study. The Appendix \ref{appendixhardcore} and \ref{appendix2djordanwigner} review the hard-core boson representation of spin and the 2d Jordan-Wigner transformation.

\section{SO(3) Majorana representation of spin} \label{secso3}

In order to introduce the SO(3) Majorana representation, we first define three Majorana fermions $\eta_{i}^{\alpha}$, $\alpha=x,y,z$ for each spin $\sigma_{i}^{\alpha}$ (throughout this section, we use $i$ and $j$ to label the position of the spin and Majorana fermion). They satisfy the following anti-commutation relations,
\begin{equation}
\{\eta_{i}^{\alpha},\eta_{j}^{\beta}\}=2\delta_{ij}\delta^{\alpha\beta}.
\end{equation} 
The SO(3) Majorana representation of spin is given by \cite{Berezin1975,Berezin1977,Tsvelik1992,tsvelikbook,Shnirman2003,Mao2003,fu2018,Shastry1997,Biswas2011}
\begin{equation}
\label{so3majoranarep}
\sigma^{x}_{i}=-i\eta_{i}^{y}\eta_{i}^{z}, \qquad \sigma^{y}_{i}=-i\eta_{i}^{z}\eta_{i}^{x},\qquad \sigma^{z}_{i}=-i\eta_{i}^{x}\eta_{i}^{y}.
\end{equation}

The three Majorana fermions $\eta_{i}^{x}, \eta_{i}^{y}$ and $\eta_{i}^{z}$ form the fundamental representation of group SO(3), corresponding to the SU(2) rotation of spin. We can define a SO(3) singlet operator $\gamma_{i}$ using the Majorana fermion operators \cite{fu2018,Shnirman2003,Mao2003,Biswas2011},
\begin{equation}
\label{so3singlet}
\gamma_{i}=-i\eta_{i}^{x}\eta_{i}^{y}\eta_{i}^{z}.
\end{equation}
The SO(3) singlet operator commutes with Majorana fermions on the same site $[\gamma_{i},\eta_{i}^{\alpha}]=0$, and it anticommutes with Majorana fermions on different sites $\{\gamma_{i},\eta_{j}^{\alpha}\}=0$, with $i\neq j$. Therefore it commutes with all spin operators, $[\gamma_{i},\sigma_{j}^{\alpha}]\equiv 0$, no matter if $i=j$ or $i\neq j$. Furthermore, it follows that the $\gamma_{i}$ operator a constant of motion because it commutes with all kinds of spin Hamiltonian \cite{fu2018,Shnirman2003,Mao2003,Biswas2011}. 

In terms of the SO(3) singlet we have another form of SO(3) Majorana representation (\ref{so3majoranarep})
\begin{equation}
\label{so3majoranaintau}
\sigma^{x}_{i}=\gamma_{i}\eta_{i}^{x}, \qquad \sigma^{y}_{i}=\gamma_{i}\eta_{i}^{y},\qquad \sigma^{z}_{i}=\gamma_{i}\eta_{i}^{z}.
\end{equation}
From this expression, we can easily see the SO(3) structure and this form has certain advantages since $\gamma$ operators are constants of motion. 

For the next step, we pair up Majorana fermions $\eta^{x}$ and $\eta^{y}$ and define complex fermion
\begin{equation}
\label{complexfermionxy}
c_{i}^{\dagger}=\frac{1}{2}(\eta_{i}^{x}+i\eta_{i}^{y}), \qquad c_{i}=\frac{1}{2}(\eta_{i}^{x}-i\eta_{i}^{y}).
\end{equation}
In terms of these complex fermions we have the spin raising and lowering operators
\begin{equation}
\label{so3ineta}
\sigma_{i}^{+}=\frac{1}{2}(\sigma_{i}^{x}+i\sigma_{i}^{y})=\eta_{i}^{z}c_{i}^{\dagger}, \quad \sigma_{i}^{-}=\frac{1}{2}(\sigma_{i}^{x}-i\sigma_{i}^{y})=c_{i}\eta_{i}^{z}.
\end{equation}
And there is another form with the SO(3) singlet,
\begin{equation}
\label{so3intau}
\sigma_{i}^{+}=\frac{1}{2}\gamma_{i}(\eta_{i}^{x}+i\eta_{i}^{y})=\gamma_{i}c_{i}^{\dagger}, \quad \sigma_{i}^{-}=\frac{1}{2}\gamma_{i}(\eta_{i}^{x}-i\eta_{i}^{y})=\gamma_{i}c_{i}.
\end{equation}
On the other hand, in terms of complex fermion (\ref{complexfermionxy}), the z component of the spin operators is written as
\begin{equation}
\label{so3sigmaz}
\sigma_{i}^{z}=2c_{i}^{\dagger}c_{i}-1=2n_{i}-1.
\end{equation}
Here and hereafter in this paper, we use $n_{i}=c_{i}^{\dagger}c_{i}$ to label the number of complex fermions. With the complex fermion, we can find a useful relation between $\gamma$ operator and the $\eta^{z}$ operators,
\begin{equation}
\label{tauandetaz}
\gamma_{i}=\sigma_{i}^{z}\eta_{i}^{z}=(2n_{i}-1)\eta_{i}^{z}=-(-1)^{n_{i}}\eta_{i}^{z},
\end{equation}
in which we have used the fact that $(-1)^{n_{i}}=(1-2n_{i})$ for the fermion number $n_{i}$ can only take two values 0 and 1. 

At this stage it is important to analyze the Hilbert space of the Majorana fermions introduced to represent the spin space. Suppose we have $N$ spin in our spin model, then the original spin Hilbert space has dimension $2^{N}$. We introduce 3 Majorana fermions to represent each spin, each Majorana fermion has Hilbert space dimension $\sqrt{2}$ \cite{Biswas2011}, thus the dimension of the Hilbert space of the Majorana fermions is $2^{\frac{3N}{2}}$. The dimension of the Majorana fermion Hilbert space is $2^{\frac{N}{2}}$ larger than the spin Hilbert space \cite{fu2018,Biswas2011}.

In Ref. \onlinecite{fu2018} and Ref. \onlinecite{Biswas2011} it was shown that one way to eliminate the additional dimension is to pair up the $N$ spin sites, forming $\frac{N}{2}$ pairs. For each pair $\langle ij\rangle$ we take the operator $\gamma_{i}\gamma_{j}$ and fix its value to be $+i$ (or equivalently $-i$). Since these operators commute with each other and they all commute with the Hamiltonian, their eigenvalues are good quantum numbers and fixing them eliminates the extra $2^{\frac{N}{2}}$ dimensions. To see this we note that the $\gamma_{i}\gamma_{j}$ operators for all the pairs are $Z_{2}$ variables whose eigenvalue can only take $\pm i$, and that the total number of constraints we apply is $\frac{N}{2}$. In Sec. \ref{sec1djordanwigner} we will compare the SO(3) Majorana representation and the one-dimensional (1d) Jordan-Wigner transformation. From this we will see another way to eliminate the extra degrees of freedom in the Hilbert space. We will also discuss the origin of a $Z_{2}$ redundancy that always appears when we apply the SO(3) Majorana representation with $\frac{N}{2}$ constraints like these.   

With these definitions at hand, one can start looking at spin Hamiltonians. Here, for the convenience of the discussion in later sections, we use the SO(3) Majorana representation to transform the Hamiltonian of the XXZ Heisenberg model, namely,
\begin{equation}
\label{XXZHeisenbergH}
\mathcal{H}_{XXZ}=\sum_{ij}J_{z}\sigma_{i}^{z}\sigma_{j}^{z}+J_{\pm}(\sigma_{i}^{+}\sigma_{j}^{-}+h.c.).
\end{equation}
First, using (\ref{so3sigmaz}), we have the $J_{z}$ term
\begin{equation}
\label{JzpartofXXZ}
J_{z}\sigma_{i}^{z}\sigma_{j}^{z}=J_{z}(2n_{i}-1)(2n_{j}-1),
\end{equation}
which is a fermion density-density interaction. The XY part of the Hamiltonian is what we will focus on. With (\ref{so3ineta}) and (\ref{so3intau}) we can rewrite the bilinear spin interaction terms of the XY Hamiltonian as the following
\begin{equation}
\label{XYinetaz}
\sigma_{i}^{+}\sigma_{j}^{-}+\sigma_{i}^{-}\sigma_{j}^{+}=\eta_{i}^{z}\eta_{j}^{z}(c_{i}c_{j}^{\dagger}+c_{i}^{\dagger}c_{j}),
\end{equation}
and in terms of $\gamma$ operators we have
\begin{equation}
\label{XYinso3}
\sigma_{i}^{+}\sigma_{j}^{-}+\sigma_{i}^{-}\sigma_{j}^{+}=-(\gamma_{i}\gamma_{j})(c_{i}^{\dagger}c_{j}+c_{i}c_{j}^{\dagger}).
\end{equation}
Therefore we see that the under the SO(3) Majorana representation, the XY Hamiltonian is transformed into a hopping of complex fermions (defined in Eq. (\ref{complexfermionxy})) coupled to link variables defined in terms of Majorana fermion $\eta^{z}$ or the SO(3) singlet operator $\gamma$. With these results, we move on to discuss the relationship between the SO(3) Majorana representation and the Jordan-Wigner transformation in one and two dimensions.

\section{Relation between the SO(3) Majorana representation of spin and the one-dimensional Jordan-Wigner Transformation} \label{sec1djordanwigner}

The Jordan-Wigner transformation defines a non-local transformation of a one-dimensional spin chain \cite{jordan1928,lieb1961,fradkinbook}. As we will see below, the non-local nature of the Jordan-Wigner transformation makes it directly comparable to the SO(3) Majorana representation of spin. Here, we emphasize again that although the SO(3) Majorana representation acts as a local transformation between spin and Majorana fermions (see (\ref{so3majoranarep}) and (\ref{so3majoranaintau})), the Hilbert space of the Majorana fermions can be defined only by pairing up the Majorana fermions non-locally because we have an odd number of Majorana fermions per site. 

We start by considering a one-dimensional spin chain. For a spin chain, it is convenient to label the position of the spin sites as $i=1,2,3,...,N$ (throughout this section, we use $N$ to denote the total number of spins in the spin chain). The Jordan-Wigner transformation in 1d takes the form \cite{jordan1928,lieb1961,fradkinbook}
\begin{equation}
\label{jordanwigner1d}
\sigma^{+}_{i}=c^{\dagger}_{i}e^{i\pi \sum_{j=1}^{i-1}c_{j}^{\dagger}c_{j}},\qquad \sigma^{-}_{i}=c_{i}e^{-i\pi \sum_{j=1}^{i-1}c_{j}^{\dagger}c_{j}}.
\end{equation}
Comparing with (\ref{so3ineta}) and (\ref{so3intau}) one notice that the Majorana fermion operator $\eta^{z}$ and $\gamma$ in the SO(3) Majorana representation acts like the quasi-infinite string operator in (\ref{jordanwigner1d}). This provides a guidance for us to discuss the correspondence between the SO(3) Majorana representation and the JW transformation in the 1d spin chain.

In order to establish the correspondence, our first task is to eliminate the extra dimensions (which is $2^{\frac{N}{2}}$ as discussed above) in the Majorana Hilbert space. This is achieved by enforcing a number of constraints on the Majorana Hilbert space. Specifically, let us denote the many-body physical space of $\eta^{x}_{i}$ and $\eta_{i}^{y}$ Majorana fermions as $\mathbb{H}_{\eta^{x}}$ and $\mathbb{H}_{\eta^{y}}$. The product space $\mathbb{H}_{xy}=\mathbb{H}_{\eta^{x}}\otimes\mathbb{H}_{\eta^{y}}$ has dimension $2^{N}$. If we assign a single state from the many-body physical space of $\eta^{z}$, which we call $\mathbb{H}_{\eta^{z}}$, to each and every state in $\mathbb{H}_{xy}$, the resulting space $\mathbb{H}_{xy}'$, which is a subspace of $\mathbb{H}_{xyz}=\mathbb{H}_{\eta^{x}}\otimes\mathbb{H}_{\eta^{y}}\otimes\mathbb{H}_{\eta^{z}}$, will still have dimension $2^{N}$ and it is what we want. To make such assignment, one need some conditions or constraints. Here, following the definition of the SO(3) Majorana representation in Sec. \ref{secso3}, we apply the following conditions 
\begin{equation}
\label{mappingso3jordan1}
\eta_{2k-1}^{z}\sim (-1)^{\sum_{j=1}^{2k-1}n_{j}}, \qquad i\eta_{2k}^{z}\sim (-1)^{\sum_{j=1}^{2k}n_{j}},
\end{equation}
in which $k=1,2,3,...$ and we use $n_{j}=c_{j}^{\dagger}c_{j}$ to denote the number operator of the complex fermion $c_{j}$ defined in (\ref{complexfermionxy}).  Using (\ref{tauandetaz}), we see that the mapping (\ref{mappingso3jordan1}) corresponds to $\gamma_{2k-1}\sim -(-1)^{\sum_{j=1}^{2k-2}n_{j}}$ and $i\gamma_{2k}\sim -(-1)^{\sum_{j=1}^{2k-1}n_{j}}$. Comparing the definition of the SO(3) Majorana representation in (\ref{so3intau}) and the JW transformation (\ref{jordanwigner1d}), these conditions mean that the $c$ fermion in SO(3) Majorana representation corresponds to the fermion in Jordan-Wigner transformation up to some extra phases; at sites $2k-1$, the phase is $-1$, at sites $2k$, the phase is $-i$. These extra phases have no influence on the definition of fermion number. 

One may argue that the mapping in (\ref{mappingso3jordan1}) is not mathematically rigorous because the left-hand-side is fermionic while the right-hand-side is bosonic. However, such discrepancy is not physical, all physical quantities must be functions of spin operators which come with the complex fermion operator $c$. After mutiplying the complex fermion operator, all the commutation relation is restored within the one-dimensional spin chain geometry. In this sense there is no problem in (\ref{mappingso3jordan1}). In physical applications, it is clearer to define an equivalent form of the mapping. Using the relation between $\gamma$ and $\eta^{z}$ (\ref{tauandetaz}) we see that the mapping (\ref{mappingso3jordan1}) is equivalent to the following up to a global $Z_{2}$ degree of freedom, 
\begin{equation}
\label{mappingso3jordan2}
i\eta_{2k-1}^{z}\eta_{2k}^{z}\sim (-1)^{n_{2k}},
\end{equation}
\begin{equation}
\label{mappingso3jordan3}
i\gamma_{2k}\gamma_{2k+1}\sim (-1)^{n_{2k}},
\end{equation}
in which $k=1,2,3,...$. Here in the mapping (\ref{mappingso3jordan2}) and (\ref{mappingso3jordan3}), both sides are bosonic operators.  

To give a physical intepretation of the conditions (\ref{mappingso3jordan2}) and (\ref{mappingso3jordan3}), we pair up the Majorana fermion $\eta_{2k-1}^{z}$ and $\eta_{2k}^{z}$ and define complex fermion $d_{2k}=\frac{1}{2}(\eta_{2k-1}^{z}-i\eta_{2k}^{z})$ whose locations are defined on sites with an even number. Since we have $i\eta_{2k-1}^{z}\eta_{2k}^{z}=1-2d_{2k}^{\dagger}d_{2k}=(-1)^{n_{d_{2k}}}$, the mapping (\ref{mappingso3jordan2}) means that the number of $d$ fermion on site $2k$ is equal to the number of $c$ fermion (formed by $\eta^{x}$ and $\eta^{y}$ Majorana fermion) on site $2k$. There are $\frac{N}{2}$ $d$ fermions and so there are $\frac{N}{2}$ such conditions, which fix the state of the $d$ fermion once the state of $c$ fermion is defined. In this way, we have assigned a state in $\mathbb{H}_{\eta^{z}}$ to each and every state in $\mathbb{H}_{xy}$. Therefore the $\frac{N}{2}$ conditions in (\ref{mappingso3jordan2}) are already sufficient to fix the dimension of the Hilbert space to be that of the spin space. But there is another $\frac{N}{2}$ constaints which take the form as (\ref{mappingso3jordan3}). At first sight, the conditions (\ref{mappingso3jordan2}) and (\ref{mappingso3jordan3}) seem to be overcomplete.

To remedy this, we note that the extra $\frac{N}{2}$ constraints in (\ref{mappingso3jordan3}) actually fix the remaining $Z_{2}$ gauge redundancy of the Majorana fermions. The SO(3) Majorana fermion representation involves bilinear form of Majorana fermions. Under the sign flip $\eta^{\alpha}\rightarrow -\eta^{\alpha}$ with $\alpha=x,y,z$, the original spin operator in (\ref{so3majoranarep}) is invariant. After enforcing the conditions (\ref{mappingso3jordan2}) there is still some $Z_{2}$ redundancy left. To see this, we note that the spin operator in the representation (\ref{so3majoranarep}) and the conditions in (\ref{mappingso3jordan2}) are invariant under simultaneous sign flipping $\eta_{2k-1}^{\alpha}\rightarrow -\eta_{2k-1}^{\alpha}$ and $\eta_{2k}^{\alpha}\rightarrow -\eta_{2k}^{\alpha}$ with $k$ being an arbitrary integer and $\alpha=x,y,z$. Although the dimension of the Hilbert space is $2^{N}$ once the first $\frac{N}{2}$ constaints in (\ref{mappingso3jordan2}) are enforced, the Hilbert space is still $2^{\frac{N}{2}}$ times larger than the spin space. Due to the remaining $Z_{2}$ gauge redundancy, for each state in the spin space, there are $2^{\frac{N}{2}}$ states in the corresponding Majorana Hilbert space $\mathbb{H}_{xy}'$. Once the other $\frac{N}{2}$ gauge fixing constaints of (\ref{mappingso3jordan3}) are enforced, the remaining gauge redundancy is eliminated. 

Therefore, the mapping (\ref{mappingso3jordan1}) or (\ref{mappingso3jordan2}) and (\ref{mappingso3jordan3}) give a correspondence between the SO(3) Majorana representation and the 1d Jordan-Wigner transformation. The SO(3) Majorana representation (Eq. (\ref{so3ineta}) and Eq. (\ref{so3intau})) with some proper constraints (Eq. (\ref{mappingso3jordan2}) and Eq. (\ref{mappingso3jordan3})) to fix the extra degrees of freedom will lead us to the same form as the 1d Jordan-Wigner transformation in (\ref{jordanwigner1d}). Throughout our discussion, we make no reference to the specific form of the spin Hamiltonian of the spin chain, thus the correspondence is between the two spin representations and can be applied to any one-dimensional spin Hamiltonian. On the other hand, the constraints in (\ref{mappingso3jordan2}) and (\ref{mappingso3jordan3}) take different form from the constrains that are previously discussed \cite{fu2018, Biswas2011}, in which we pair up sites and demand that for each pair $\langle ij\rangle$, $\gamma_{i}\gamma_{j}=-i$. In general there are multiple ways to fix the extra degrees of freedom in the SO(3) Majorana representation. Different fixing will lead to different forms of the resulting theory.   

It is important to emphasize that the complete elimination of extra degree of freedom in Majorana Hilbert space is only achievable in one-dimensional spin chain. In 1d spin chain, after we pair up sites and enforce the first $\frac{N}{2}$ constraints to eliminate the extra dimension of the Majorana Hilbert space (like the ones in (\ref{mappingso3jordan2})), the rest of the link variables decouple and allow us to fix the extra $Z_{2}$ redundancy by introducing another set of constraints (like (\ref{mappingso3jordan3})). In higher dimensional space, the number of links connecting to each site is larger than two, it is generally impossible to define the second set of constraints. Without the extra gauge-fixing constraints like in (\ref{mappingso3jordan3}), the original spin model is always mapped to some $Z_{2}$ gauge theory with complex fermion as its matter field. In Sec. \ref{secxyhoneycomb} and Sec. \ref{seccompass}, we study two spin models using the SO(3) Majorana representation, namely the quantum XY model on honeycomb lattice and the $90^{\circ}$ compass model on the square lattice. We explicitly show that, if only $\frac{N}{2}$ constraints are enforced, both models can be mapped into some non-trivial $Z_{2}$ gauge theory. In our discussion, to get the $\frac{N}{2}$ constraints, we pair up sites of the lattice and demand that for each pair $\langle ij\rangle$, the product of the SO(3) singlets $\gamma_{i}\gamma_{j}=i$ (or $-i$). Due to the fact that all $\gamma_{i}\gamma_{j}$ commute with the spin Hamiltonian, we are able to transform these constraints into the form of standard Gauss law constraints in $Z_{2}$ gauge theory \cite{Kogut1979,fradkinbook}, which commute with the $Z_{2}$ Hamiltonian by construction. 

As another example, in Ref. \onlinecite{fu2018}, it is shown that the Kitaev honeycomb model\cite{Kitaev2006} can be solved using SO(3) Majorana representation, the resulting solution takes the form of a $Z_{2}$ lattice gauge theory with standard Gauss law constraint. In other words, in the Kitaev model, due to the unique form of the Hamiltonian and the lattice geometry, it is possible to fix the $Z_{2}$ gauge without introducing any approximation. In this sense, the Kitaev model on 2d honeycomb lattice behaves like the 1d spin chain. The models we are considering in Sec. \ref{secxyhoneycomb} and Sec. \ref{seccompass} do not have such property.

\section{Relation between the SO(3) Majorana representation of spin and the two-dimensional Jordan-Wigner Transformation} \label{sec2djordanwigner}

There is a direct generalization of the Jordan-Wigner transformation to two-dimensional (2d) space with the aid of Chern-Simons gauge theory \cite{fradkin1989,lopez1994,kumar2014,wang1991,Ambjorn1989,Azzouz1993}. The two-dimensional Jordan-Wigner transformation starts with the hard-core boson representation of spin \cite{fradkinbook,fradkin1989} (see Appendix \ref{appendixhardcore} for a review), with the U(1) Chern-Simons term, the statistics of the hard-core boson can be changed to fermionic. More generally, the statistics of particles in (2+1)d spacetime are not just bosonic and fermionic \cite{wilczek1982}, particles in (2+1)d with exotic statistics are called {\it anyons} \cite{fradkinbook,Kitaev2006,Kitaev2003,nayak2008}. The 2d Jordan-Wigner transformation maps the spin operator to a complex fermion attached to a half-infinite string operator of gauge field \cite{fradkin1989,lopez1994,kumar2014}. For a lattice spin model, applying the 2d Jordan-Wigner transformation (or Chern-Simons JW transformation) requires proper definition of U(1) Chern-Simons gauge theory on a lattice \cite{lopez1994,kumar2014,sun2015,eliezer1992,eliezer92}. It is proved that the lattice Chern-Simons theory can only be defined on 2d lattices which have a one-to-one mapping between sites and plaquettes \cite{sun2015}. On 2d lattices with such property, the 2d Jordan-Wigner transformation maps a quantum XY model, whose Hamiltonian is given by
\begin{equation}
\label{quantumxymodel}
\mathcal{H}_{XY}=\sum_{ij}J_{\pm}(\sigma_{i}^{+}\sigma_{j}^{-}+\text{h.c.}),
\end{equation}
into a system of complex fermion $c_{i}$ defined on lattice sites interacting with Chern-Simons gauge field $A_{ij}$ defined on lattice bonds $\langle ij\rangle$ \cite{fradkin1989,fradkinbook},
\begin{equation}
\label{XYincs}
\mathcal{H}_{XY}=\sum_{ij}J_{\pm}c_{i}^{\dagger}e^{iA_{ij}}c_{j}+\text{h.c.}.
\end{equation} 
In Appendix \ref{appendixhardcore} and \ref{appendix2djordanwigner}, we give brief review of the hard-core boson representation of spin and the 2d Jordan-Wigner transformation using Chern-Simons terms. To lay foundation of the discussion on the relationship between SO(3) Majorana representation and the 2d Jordan-Wigner transformation, we start with a review of the basics of the Chern-Simons gauge theory and the lattice Chern-Simons theory, following Ref. \onlinecite{fradkinbook,sun2015}.

\subsection{Basics of U(1) Chern-Simons gauge theory}

The definition of Chern-Simons (CS) term relies on the existance of the total antisymmetrized tensor $\epsilon_{\mu\nu\rho}$ in (2+1)-dimensions. The definition of U(1) CS action with interaction with matter current is given by
\begin{equation}
\label{chernsimonsaction}
S_{CS}+S_{int}=\int d^{3}x (\frac{k}{4\pi}\epsilon^{\mu\nu\rho}\mathcal{A}_{\mu}\partial_{\nu}\mathcal{A}_{\rho}-J^{\mu}\mathcal{A}_{\mu}),
\end{equation} 
in which $\mathcal{A}_{\mu}$ is the Chern-Simons gauge field and matter current is given by $J^{\mu}$, all the indices $\mu,\nu,\rho=0,1,2$. Throughout this paper, we use $\mathcal{A}$ to label Chern-Simons gauge field in continuum and use $A$ to denote Chern-Simons gauge field on a lattice. The pure Chern-Simons term 
\begin{equation}
\label{purechernsimons}
S_{CS}=\frac{k}{4\pi}\int d^{3}x \epsilon^{\mu\nu\rho}\mathcal{A}_{\mu}\partial_{\nu}\mathcal{A}_{\rho}.
\end{equation}
is gauge invariant under local gauge transformation. In particular, under gauge transformation $\mathcal{A}_{\mu}\rightarrow\mathcal{A}_{\mu}-\partial_{\mu}\phi$, the action change to 
\begin{equation}
S_{CS}\rightarrow S_{CS}-\frac{k}{4\pi}\int d^{3}x \partial_{\mu}(\epsilon^{\mu\nu\rho}\phi\partial_{\nu}\mathcal{A}_{\rho}),
\end{equation} 
which vanishes because it is a total derivative. In the prefactor $\frac{k}{4\pi}$, the $k$ is called the {\it level} of the Chern-Simons theory, it can be proved that $k$ can only take integer values under the requirement that the Chern-Simons term (\ref{purechernsimons}) is gauge invariant in finite temperature \cite{fradkinbook}.

The time component of $\mathcal{A}$ does not have any dynamics, to see this we have to write the action (\ref{chernsimonsaction}) in the following way \cite{fradkinbook}
\begin{equation}
S_{t}=\int d^{3}x [(\frac{k}{2\pi}\mathcal{A}_{0}\mathcal{B}-J_{0}\mathcal{A}_{0})-\frac{k}{4\pi}\epsilon_{ij}\mathcal{A}_{i}\partial_{t}\mathcal{A}_{j}-J_{i}\mathcal{A}_{i}],
\end{equation}
in which magnetic field $\mathcal{B}$ is defined by $\mathcal{B}=\epsilon_{ij}\partial_{i}\mathcal{A}_{j}$. Upon intergrating out $\mathcal{A}_{0}$ in the path integral, we have the constraint that $\frac{k}{2\pi}\mathcal{B}-J_{0}=0$. In the canonical formulism, it should be understood as the operator on the left hand side acting on the physical states gives zero \cite{fradkinbook,sun2015}, i.e. 
\begin{equation}
\label{fluxattachmentconstraint}
[\frac{k}{2\pi}\mathcal{B}(\boldsymbol{x})-J_{0}(\boldsymbol{x})]|\text{Phys}\rangle=0.
\end{equation}
This is a requirement that the charge carried by the complex fermion $c$ must come with a magnetic flux. Due to the Aharonov-Bohm effect, the attachment of magnetic flux to charged particles results in exotic statistics of particles \cite{fradkinbook,nayak2008}. 

The CS term (\ref{purechernsimons}) has an important property, the canonical momentum conjugate to the gauge field is the gauge field itself. This results in the following non-trivial commutation relation
\begin{equation}
\label{comutationrelation}
[\mathcal{A}_{i}(\boldsymbol{x}),\mathcal{A}_{j}(\boldsymbol{y})]=i\frac{2\pi}{k}\epsilon_{ij}\delta(\boldsymbol{x}-\boldsymbol{y}).
\end{equation}
On the other hand, this property also results in the fact that the Hamiltonian of the pure Chern-Simons term (\ref{purechernsimons}) vanishes $\mathcal{H}_{CS}=0$. 

The line integral of gauge field plays important roles in gauge theories, the commutation relation (\ref{comutationrelation}) results in non-trivial commutation between line integrals. For two arbitrary lines $\mathcal{C}$ and $\mathcal{C}'$ (with directions defined) we have
\begin{equation}
\bigg[\int_{\mathcal{C}}\mathcal{A},\int_{\mathcal{C}'}\mathcal{A}\bigg]=i\frac{2\pi}{k}\nu[\mathcal{C},\mathcal{C}'],
\end{equation}
in which $\nu[\mathcal{C},\mathcal{C}']$ is the number of oriented intersections between two lines \cite{sun2015}. If $\mathcal{C}$ and $\mathcal{C}'$ are closed loops, $\nu[\mathcal{C},\mathcal{C}']$ is topologically invariant; besides, if any of $\mathcal{C}$ and $\mathcal{C}'$ can be contracted into a point, $\nu[\mathcal{C},\mathcal{C}']=0$. The line integral of gauge field can be used to construct the {\it Wilson line} operators and the {\it Wilson loop} operators. In general the Wilson line operator is defined as $W_{L}=\exp(i\int_{L}\mathcal{A}\cdot d\boldsymbol{x})$. 

Now we explore the non-trivial commutation relations between Wilson lines in Chern-Simons gauge theory. To do this, it turns out that the Baker-Hausdorff-Campbell (BHC) formula is useful; it states that for any operators X and Y, if commutator $[X,Y]$ is a number then we have
\begin{equation}
\label{BHC}
e^{X}e^{Y}=e^{X+Y+\frac{1}{2}[X,Y]}=e^{Y}e^{X}e^{[X,Y]}.
\end{equation}
For two lines $\mathcal{C}$ and $\mathcal{C}'$, we define the Wilson line operators 
\begin{equation}
W_{\mathcal{C}}=e^{i\int_{\mathcal{C}}\mathcal{A}}, \qquad W_{\mathcal{C}'}=e^{i\int_{\mathcal{C}'}\mathcal{A}}.
\end{equation}
Using the BHC formula, we have
\begin{equation}
W_{\mathcal{C}}W_{\mathcal{C}'}=W_{\mathcal{C}'}W_{\mathcal{C}}e^{-[\int_{\mathcal{C}}\mathcal{A},\int_{\mathcal{C}'}\mathcal{A}]}=W_{\mathcal{C}'}W_{\mathcal{C}}e^{-i\frac{2\pi}{k}\nu[\mathcal{C},\mathcal{C}']}.
\end{equation}
Now we focus on some special situations. In the 2d Jordan-Wigner transformation we take the level $k=1$ \cite{fradkin1989,fradkinbook} (see Appendix \ref{appendix2djordanwigner} for details), the corresponding CS theory is called the $\text{U(1)}_{1}$ Chern-Simons theory. In the $\text{U(1)}_{1}$ Chern-Simons theory we have $e^{-i\frac{2\pi}{k}\nu[\mathcal{C},\mathcal{C}']}=1$, therefore
\begin{equation}
[W_{\mathcal{C}}, W_{\mathcal{C}'}]=0.
\end{equation} 
This means that the Wilson lines in the $\text{U(1)}_{1}$ Chern-Simons gauge theory all commute with each other. This is the result for Chern-Simons theory in the continuum. The lattice version of the Chern-Simons theory has different results for Wilson lines \cite{sun2015}, which we will discuss in the next section.

\subsection{Lattice U(1) Chern Simons gauge theory}

The lattice discretization of the U(1) Chern-Simons gauge theory has been discussed on square lattice \cite{eliezer1992,eliezer92,lopez1994} and kagome lattice \cite{kumar2014}. A general discussion on the conditions for lattice Chern-Simons theory has also been done \cite{sun2015}. Here, we follow Ref. \onlinecite{sun2015} and give a brief review of some general results of lattice U(1) Chern-Simons theory, and we will focus on the situation where the level $k=1$. 

From the standard way to define lattice gauge theories \cite{Kogut1979}, we place the paricle operators on the sites of the lattice and the gauge field operators on the bonds of the lattice. To discretize the U(1) CS theory (\ref{purechernsimons}) on a lattice, it is proved that a key condition is that there is a one-to-one mapping between sites and plaquettes of the lattice. If a graph or lattice has such mapping, one can find a way to pair up the sites and plaquettes. Once the pairing is determined, the lattice CS theory will attach the gauge flux in the plaquette to the particle defined on the corresponding site. For any given 2d lattice, the three types of elements are sites (or vertices), labelled by $v$; bonds (or edges), labelled by $e$; and plaquettes (or faces) labelled by $f$. For a lattice with one-to-one correspondence between sites and plaquettes, we have the action of the lattice CS theory,
\begin{equation}
\label{latticecs}
S_{CS}=\frac{k}{2\pi}\int dt \sum_{v,f,e,e'}\bigg[ A_{v}M_{v,f}\Phi_{f}-\frac{1}{2}A_{e}K_{e,e'}\dot{A}_{e'}\bigg],
\end{equation}
in which the sum is over all sites, faces and edges of the lattice. Specifically, the flux operator $\Phi_{f}$ is defined by $\Phi_{f}=\sum_{e}\xi_{f,e}A_{e}$, in which $\xi_{f,e}=\pm 1$ if and only if $e$ is an edge of face $f$, otherwise $\xi_{f,e}=0$. The sign of $\xi_{f,e}$ is determined by the orientation of the bond. The $\Phi_{f}$ defined in this way is the lattice version of the flux. Also, in (\ref{latticecs}) the $M_{v,f}$ and $K_{e,e'}$ are two matrices. In particular $M_{v,f}$ picks up the site that is paired up with each face, its element is non-zero if and only if $v$ is paired up with $f$; the $K_{e,e'}$ matrix is defined in the following way:
\begin{eqnarray}
\begin{aligned}
&K_{e,e'}=\pm \frac{1}{2} \quad \text{if $e$ and $e'$ belong to the same face},\\ 
&K_{e,e'}=0 \quad \text{for all other cases}.
\end{aligned}
\end{eqnarray}
The sign of non-vanishing elements of $K_{e,e'}$ is determined by the orientation of the bonds and their relative positions in the face, the details of which is not important for our purpose (see Ref. \onlinecite{sun2015} for a detailed description). 

The gauge transformation in the lattice is defined by 
\begin{equation}
A_{v}\rightarrow A_{v}-\partial_{t}\tilde{\phi}_{v}, \qquad A_{e}\rightarrow A_{e}-D_{v,e}\tilde{\phi}_{v}, 
\end{equation}
in which $\tilde{\phi}_{v}$ is an arbitrary real function defined on the sites and $D_{v,e}=\pm 1$ if and only if $v$ is one of the end points of edge $e$, otherwise it is zero. As defined above, $\xi_{f,e}$ represents a lattice curl and $D_{v,e}$ represents a lattice gradient. It can be shown that the key condition for the lattice theory to be gauge invariant is that \cite{sun2015}
\begin{equation}
\sum_{f}M_{v,f}\xi_{f,e}=\sum_{e'}K_{e,e'}D_{v,e'}.
\end{equation}
It can be proved that this condition is indeed satisfied by the construction described above. 

One key property for the lattice satisfying the one-to-one correspondence between sites and faces is the existance of a dual lattice. To get the dual lattice, one simply reverses the definition of face and vertices. We put a vertex $v^{*}$ in each face of the original graph and connect two $v^{*}$ vertices if in the original graph the two faces share an edge, and thus we get the dual edge $e^{*}$. Obviously, we have the duality of each element as $v^{*}=f$, $e^{*}=e$ etc \cite{sun2015}. In the dual lattice the dual Chern-Simons theory can be defined according to (\ref{latticecs}). The $K_{e,e'}$ matrix in the dual theory becomes $K^{*}_{e^{*},e^{'*}}$. Due to the correpondence between edges $e$ and $e^{*}$, this can also be denoted as $K^{*}_{e,e'}$, its definition in the original edge indicies reads
\begin{eqnarray}
\label{Kstarmatrix}
\begin{aligned}
&K^{*}_{e,e'}=\pm \frac{1}{2} \quad \text{if e and $e'$ share a vertex}, \\
&K^{*}_{e,e'}=0 \quad \text{otherwise}.
\end{aligned}
\end{eqnarray}
It can be shown that the $K^{*}$ matrix is actually related to the inverse of the $K$ matrix,
\begin{equation}
K^{*}=-K^{-1}.
\end{equation}
so that the $K_{e,e'}$ matrix is non-singular \cite{sun2015}. 

In the canonical formulism, the commutator between gauge field on edges follows directly from the Lagrangian, which is the integrand in (\ref{latticecs}),
\begin{equation}
[A_{e},\frac{k}{2\pi}K_{e',e''}A_{e''}]=i\delta_{e,e'}.
\end{equation}
Since the $K$ matrix can be inverted, we have
\begin{equation}
\label{commutationonlattice}
[A_{e},A_{e'}]=-\frac{2\pi i}{k}K^{-1}_{e,e'}.
\end{equation}
The flux attachment on the lattice work similarly as the continous case, we place charge density $J^{0}_{v}$ on each vertex $v$ and couple it to $A_{v}$. We thus have the constraint
\begin{equation}
\label{fluxattachonlattice}
[\frac{k}{2\pi}M_{v,f}\Phi_{f}-J^{0}_{v}]|\text{Phys}\rangle=0.
\end{equation}
With these results at hand, we have a consistent theory of lattice Chern-Simons gauge theory. 

In order to discuss the relationship between the SO(3) Majorana representation of spin and the 2d Chern-Simons Jordan-Wigner transformation, we need one more element, which is the compactification of the lattice U(1) Chern-Simons gauge theory. 

\subsection{Compactification of U(1) Chern-Simons gauge theory on a lattice}

As with other types of lattice gauge theories, the gauge field in the lattice Chern-Simons theory couples to the matter field by a Wilson line \cite{Kogut1979,fradkinbook,lopez1994,kumar2014}, 
\begin{equation}
\mathcal{H}\sim c_{i}^{\dagger}e^{iA_{ij}}c_{j}+\text{h.c.},
\end{equation}
in which $A_{ij}$ is the lattice gauge field defined on the bond $\langle ij\rangle$. Throughout this section, we interchangably use $\langle ij\rangle$ (contains the start point and the end point of the bond) and $e$ to label the bonds of the lattice. We note that the gauge field on the bond $A_{e}$ actually corresponds to line integral of gauge field $\mathcal{A}_{\mu}$ in the continuous theory. The Wilson line on each bond $e$ takes the form of $W_{e}=e^{iA_{e}}$, we call them the {\it Wilson link variables} (or Wilson links). The Wilson links are invariant under the the addition of integer multiples of $2\pi$ to the gauge field on the link. This requires that the lattice Chern-Simons gauge field is defined in a compact manifold. The compactification of the gauge field  $A_{e}$ means that $A_{e}$ and $A_{e}+2n\pi$ are always equavilent when $n$ is an integer. In other words, we have
\begin{equation}
\label{compactificationrequirement}
A_{e}+2\pi\equiv A_{e}.
\end{equation} 

From previous discussion, we have that the commutator of the gauge field on a lattice is given by (\ref{commutationonlattice}). It follows from (\ref{Kstarmatrix}) that
\begin{equation}
[A_{e},A_{e'}]=-\frac{2\pi i}{k}K_{e,e'}^{-1}=\frac{2\pi i}{k}(\pm\frac{1}{2})=\pm \frac{i\pi }{k},
\end{equation}
{\it when $e$ and $e'$ share a vertex}. For 2d Jordan-Wigner transformation, we are taking the level $k=1$, so we have
\begin{equation}
\label{gaugefieldcommutatoronlattice}
[A_{e},A_{e'}]=\left\{ \right.\begin{array}{cc}
\pm i\pi & \text{if $e$ and $e'$ share a vertex},\\
0 & \text{otherwise}. 
\end{array}
\end{equation}
Now we suppose that $[A_{e},A_{e'}]=i\pi$, which means that 
\begin{equation}
[A_{e},\frac{A_{e'}}{\pi}]=i.
\end{equation}
Then we have operator identity
\begin{equation}
e^{i\frac{A_{e'}}{\pi}\theta}A_{e}e^{-i\frac{A_{e'}}{\pi}\theta}=A_{e}+\theta.
\end{equation}
Specifically when $\theta=2\pi$, we have the following, using condition (\ref{compactificationrequirement}),
\begin{equation}
\label{compactificationrelation}
e^{2iA_{e'}}A_{e}e^{-2iA_{e'}}=A_{e}+2\pi\equiv A_{e}.
\end{equation}
To ensure this is an identity for all $A_{e}$ we have to require that $e^{2iA_{e'}}=C$, where $C$ is a constant. Eq. (\ref{compactificationrelation}) implies that $|C|^{2}=1$, which means
\begin{equation}
\label{compactificationconsequence}
e^{2iA_{e'}}=e^{i\phi_{e'}},
\end{equation}
in which $\phi_{e'}$ is a constant phase defined on bond $e'$.

On the other hand, if $[A_{e},A_{e'}]=-i\pi$, we have $[A_{e},-\frac{A_{e'}}{\pi}]=i$. This leads to
\begin{eqnarray}
e^{-2iA_{e'}}A_{e}e^{2iA_{e'}}=A_{e}+2\pi\equiv A_{e}.
\end{eqnarray} 
Once again we arrive at the requirement (\ref{compactificationconsequence}). In summary, to compactify gauge field defined on bond $e$, we have to require that on all the bonds that share a vertex with it, the gauge field satisfies (\ref{compactificationconsequence}). Since the lattices we are interested in are always connected, all the bonds have some other neighbouring bonds, to compactify all the gauge field, we have to require that 
\begin{equation}
\label{compactificationresult}
e^{2iA_{e}}=e^{i\phi_{e}}, \quad \text{for all the links $e$ of the lattice}.
\end{equation}
In (\ref{compactificationresult}), the constant $\phi_{e}$ can vary from bond to bond. On each bond, there are multiple solutions for (\ref{compactificationresult}) for each value of the constant phase $\phi_{e}$, namely $A_{e}=\frac{1}{2}\phi_{e}+n\pi$, where $n$ is an integer. If we restrict that $0\leq \phi_{e}< 2\pi$, then there are two solutions for $A_{e}$ satisfying $0\leq A_{e}< 2\pi$, which are $A_{e}=\frac{1}{2}\phi_{e}$ and $A_{e}=\frac{1}{2}\phi_{e}+\pi$. For all values of $\phi_{e}$, we see that under the condition of compactification (\ref{compactificationrequirement}), the lattice $\text{U(1)}_{1}$ Chern-Simons gauge theory naturally breaks down to a $Z_{2}$ theory whose Wilson link can only take eigenvalues $e^{i\frac{\phi_{e}}{2}}$ or $e^{i(\frac{1}{2}\phi_{e}+\pi)}$.

There is another intepretation of this result. The commutation relation of gauge fields (\ref{gaugefieldcommutatoronlattice}) means that the Hilbert space of the lattice gauge field $A_{e}$ is not a ``coordinate space", instead, it is a phase space containing both coordinate and momentum degrees of freedom. Consistency requires that this Hilbert space (or phase space) is defined in a compact manifold with finite volume. Quantization of a phase space with finite volume always results in Hilbert space with finite dimension \cite{Tong2016}. This is the origin of the quantization of the gauge field in the lattice $\text{U(1)}_{1}$ Chern-Simons gauge theory.     

With these results, we are ready to discuss the relationship between the SO(3) Majorana representation of spin and the compactified U(1) Chern-Simons Jordan-Wigner transformation.

\subsection{SO(3) Majorana representation of spin as compactified Chern-Simons Jordan-Wigner transformation}

In the Chern-Simons Jordan-Wigner transformation of spin in 2d, any spin Hamiltonian which is bilinear in spin operators is mapped to a lattice model of fermion intereacting with Chern-Simons gauge field. In particular the XY spin Hamiltonian is mapped according to 
\begin{equation}
\label{XYinchernsimons}
\sigma_{i}^{+}\sigma_{j}^{-}+\sigma_{i}^{-}\sigma_{j}^{+}=c_{i}^{\dagger}e^{iA_{ij}}c_{j}+\text{h.c.}.
\end{equation}  
Based on the Baker-Hausdorff-Campbell formula (\ref{BHC}), the Wilson link variables on the lattice $W_{e}=e^{iA_{e}}$, or $W_{ij}=e^{iA_{ij}}$ satisfy the following relation $W_{e}W_{e'}=W_{e'}W_{e}e^{-[A_{e},A_{e'}]}$. Using the commutator (\ref{gaugefieldcommutatoronlattice}) we arrive at the commutation relations between Wilson links on the lattice,
\begin{equation}
\label{commutatorofwilsonlink}
\begin{array}{cc}
\big\{W_{e},W_{e'}\big\}=0, & \text{if $e$ and $e'$ share a vertex},\\
\big[W_{e},W_{e'}\big]=0, & \text{otherwise}.
\end{array}
\end{equation}

On the other hand, in the SO(3) Majorana representation of spin, the XY spin Hamiltonian is mapped according to Eq. (\ref{XYinso3}), the link variables are $\gamma_{i}\gamma_{j}$ on link $\langle ij\rangle$. Following the commutation relation of Majorana fermions, these link variables satisfies the following commutation relations 
\begin{equation}
\label{commutationso3link}
\begin{array}{cc}
\big\{\gamma_{i}\gamma_{j},\gamma_{j}\gamma_{k}\big\}=0, & \text{for $k\neq i$},\\
\big[\gamma_{i}\gamma_{j},\gamma_{k}\gamma_{l}\big]=0, & \text{for $k\neq i,j$ and $l\neq i,j$}.
\end{array}
\end{equation}
In other words, the link variables in the SO(3) Majorana representation of the spin model have the following commutation relation: two link variables anticommute if they share a vertex, otherwise they commute with each other. This is the same commutation relations as the Wilson links in the Chern-Simons Jordan-Wigner transformation of spin, which is given by (\ref{commutatorofwilsonlink}). Based on this similarity and compare Eq. (\ref{XYinso3}) and Eq. (\ref{XYinchernsimons}) we arrive at the following correspondence between the link variables in the SO(3) Majorana representation and the Wilson links in the lattice Chern-Simons Jordan-Wigner transformation,
\begin{equation}
\label{so3andcs}
(-\gamma_{i}\gamma_{j})\sim e^{iA_{ij}}, 
\end{equation}

The correspondence in (\ref{so3andcs}) is not complete until we analyze the eigenvalues of the link variables in both representations. According to the compactification of the gauge field in lattice $\text{U(1)}_{1}$ Chern-Simons gauge theory, its Wilson links can only take $Z_{2}$ values. Specifically, we have the Wilson links $e^{iA_{e}}$ take values $e^{i\frac{\phi_{e}}{2}}$ or $e^{i(\frac{\phi_{e}}{2}+\pi)}$ for constant $\phi_{e}$ satisfying $0\leq \phi_{e}<2\pi$. If we take $\phi_{e}\equiv\pi$ for all the bonds $e$, then the Wilson links take values $e^{iA_{e}}=\pm i$. On the other hand, for the SO(3) Majorana representation of spin we also have $(-\gamma_{i}\gamma_{j})=\pm i$. Therefore the link variables and the Wilson links in both sides of Eq. (\ref{so3andcs}) can have the same eigenvalues. To clarify the physical meaning of the condition, $\phi_{e}\equiv\pi$ or $e^{2iA_{e}}\equiv -1$ for all bonds, we point out the following intepretation. Every time the gauge field change by $2\pi$ on each bond the wavefunction (of the whole system of fermions and gauge field) goes back to itself but acquire a phase $e^{i\pi}=-1$. This phase is identified as a {\it Berry phase} \cite{berry1984,xiao2010,aharonov1987} since the gauge field is defined on a compact manifold.

In the discussion above, we have used the XXZ Heisenberg model (given by Eq. (\ref{XXZHeisenbergH})) in 2d as an example to analyze the relation between the SO(3) Majorana representation and the Chern-Simons JW transformation. From the transformation of the XY part of the Hamiltonian, we find that the complex fermions in both representations (in SO(3) Majorana, complex fermion is defined by Eq. (\ref{complexfermionxy})) are identified with each other and the Chern-Simons Wilson links are identified with the link variables in SO(3) Majorana representation. In addition, we point out that the $J_{z}$ part of the Hamiltonian under both representations are exactly the same four fermion interaction, given by Eq. (\ref{JzpartofXXZ}). These results can be directly generalized to other spin Hamiltonians in two dimensions and the correspondence we find is between two spin representations without reference to specific spin models.  

In summary of the discussion, we have the conclusion that the SO(3) Majorana representation of spin is equivalent to the Chern-Simons Jordan-Wigner transformation in two dimensions under the condition that the $\text{U(1)}_{1}$ Chern-Simons gauge field in the latter is compactified with a Berry phase $e^{i\pi}$. Such equivalence has several implications. Most importantly, from the equivalence (\ref{so3andcs}) and the commutator (\ref{commutatorofwilsonlink}), (\ref{commutationso3link}) we see that the key property of both representations is the {\it anticommuting link variables}. Previous study of the Chern-Simons JW transformation \cite{fradkin1989,lopez1994,kumar2014} uses field theoretical approach and look for the saddle point of the gauge field configuration. Such approach neglects the anticommuting nature of the neighbouring link variables. In some sense it corresponds to a mean-field treatment of the anticommuting link variables. In the SO(3) Majorana representation, previous studies \cite{Herfurth2013,Shastry1997,Biswas2011} also use mean-field approach to handle the link variables, which turns out to results in large discrepancies with the real physical states \cite{Shastry1997}. In general, there is some difficulties in the treatment of anticommuting link variables. However, as we show in the previous section, it is possible to get rid of the anti-commuting link variables in one-dimensional systems due to the unique lattice geometry. As a special two-dimensional case, in the solution of the Kitaev model using SO(3) Majorana representation \cite{fu2018} the anticommuting link variables are mapped out due to the specific form of spin Hamiltonian and lattice geometry. For general spin models in two dimensions and beyond, we do not expect such possibility.

Besides the similarities discussed previously, it is also important to note the subtleties in the correspondence between the SO(3) Majorana representation and the Chern-Simons Jordan-Wigner transformation (\ref{so3andcs}). First, the definition of the Chern-Simons Jordan-Wigner transformation is restricted to two-dimensional space in which the Chern-Simons gauge theory exists. Specific to two-dimensional space, the proper definition of the 2d Jordan-Wigner transformation requires that the lattice has a one-to-one correspondence between its sites and plaquettes \cite{sun2015}. On the contrary, the SO(3) Majorana representation can be applied in any spatial dimension and in two-dimensional space, it can be applied to any type of lattice. Moreover, due to the definition of the SO(3) singlet $\gamma$ in SO(3) Majorana representation, the fermion operators defined on site $i$ always anticommute with the link variables $\gamma_{i}\gamma_{j}$ that are connected to it. There is no such anticommuting relations in the Chern-Simons JW transformation. These discrepancies mean that the equivalence between the SO(3) Majorana representation and the 2d Jordan-Wigner transformation is {\it not mathematically rigorous}. We can understand it in the following way. Whenever the 2d Jordan-Wigner transformation can be applied to some spin model, the SO(3) Majorana representation can provide an alternative operator form for it. In general, the SO(3) Majorana representation can be applied to a broader range of models.   

On the other hand, we should also mention the limitation of the theory. In particular, we note that there should always be a Maxwell term $S_{M}=-\frac{1}{4}\int d^{3}x\mathcal{F}^{\mu\nu}\mathcal{F}_{\mu\nu}$ (in which $\mathcal{F}^{\mu\nu}$ is the standard field strength tensor for the gauge field) coming along with the pure Chern-Simons term in the total continuous action (\ref{chernsimonsaction}). The Maxwell action will make sure that the Hamiltonian is bounded from below. After including the Maxwell action the theory becomes a {\it Maxwell-Chern-Simons theory} \cite{jackiw1990,diamantini1993}. Specifically, the flux attachment constraint (\ref{fluxattachmentconstraint}) and the commutator between gauge field (\ref{comutationrelation}) are modified accordingly, including the contribution from the electric field. In the continuum limit, the Chern-Simons term will give the gauge field a mass \cite{jackiw1990}, making the interaction coming from the Maxwell term short-ranged, thus we can ignore the Maxwell part if we are only interested in long distances. However, things are different for the lattice version of the theory. Whether it is still possible to ignore the Maxwell term in the lattice Chern-Simons gauge theory is still an open question. If we include the Maxwell term in the lattice Chern-Simons theory, all the commutation relation discussed in this section will have to be modified significantly, including the compactification of gauge field. Exploration of the lattice Maxwell-Chern-Simons theory is beyond the scope of this work and left for future study.

Summarizing Sec. \ref{sec1djordanwigner}  and Sec. \ref{sec2djordanwigner}, we find that there is a correspondence between the SO(3) Majorana representation and the Jordan-Wigner transformation in both 1d and 2d under certain circumstancies. In Sec. \ref{sec1djordanwigner} we see that under the SO(3) Majorana represetation general spin models will be mapped into a $Z_{2}$ gauge theory if only $\frac{N}{2}$ ($N$ is the total number of spin in the system) fixing conditions are imposed. In Sec. \ref{sec2djordanwigner} we see the importance of anticommuting link variables in both the SO(3) Majorana representation and the Chern-Simons JW transformation. To explore the application of the SO(3) Majorana representation, we will consider two spin models, namely the quantum XY model on the honeycomb lattice and the $90^{\circ}$ compass model on the square lattice. We will map the two models into some lattice $Z_{2}$ gauge theories using the SO(3) Majorana representation. Our treatments of the two spin models is unique in that no approximation is introduced in obtaining the $Z_{2}$ gauge theory.

\section{Application of the SO(3) Majorana representation in two spin models} \label{sectwospinmodels}

\subsection{Quantum XY Model on Honeycomb Lattice} \label{secxyhoneycomb}

\begin{figure}
\includegraphics[width=0.4\textwidth]{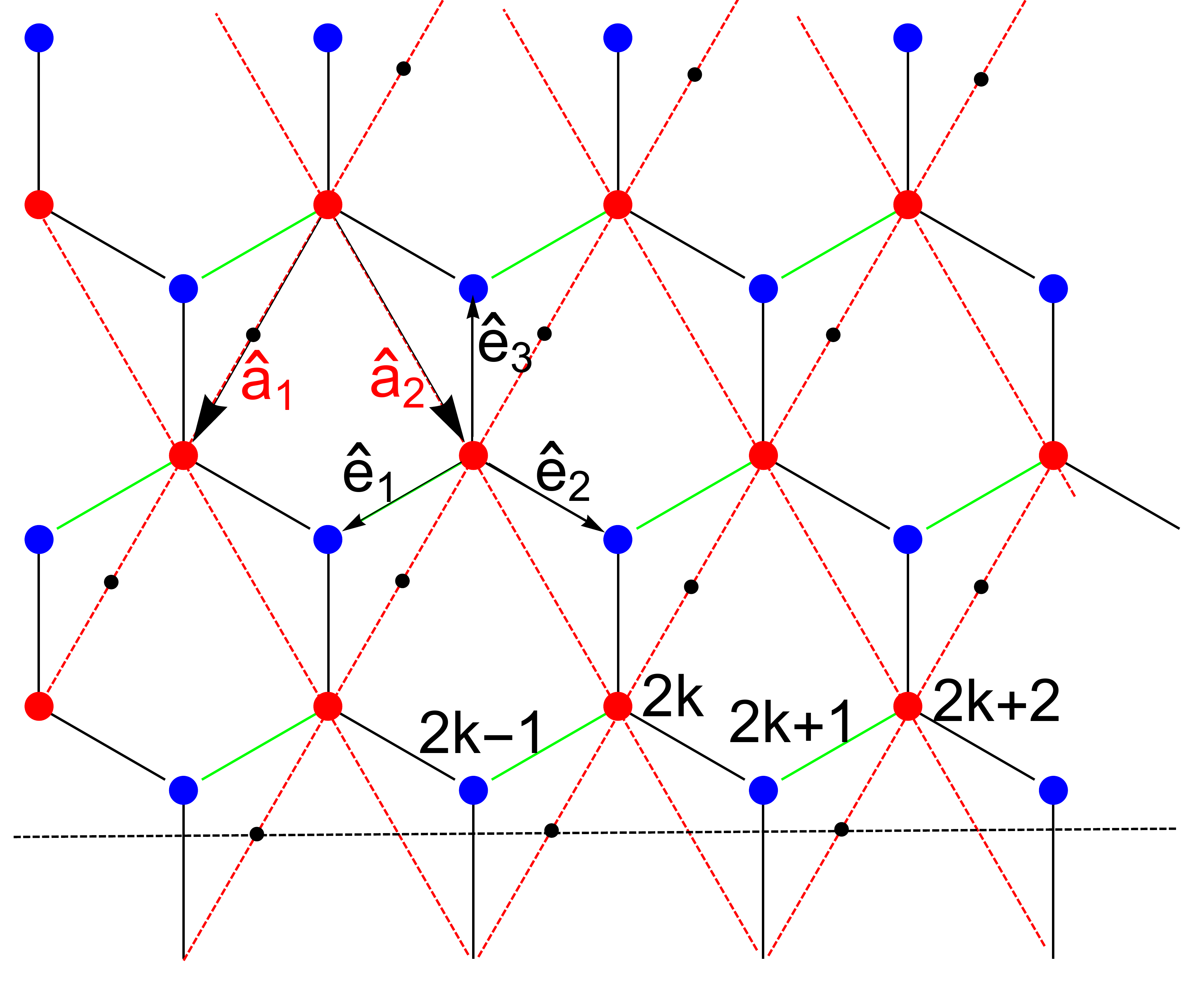}
\caption{The honeycomb lattice and the diamond lattice. The original spins in the quantum XY model are defined on the sites of the honeycomb lattice, the three types of bonds are labelled by vectors $\hat{e}_{1},\hat{e}_{2}$ and $\hat{e}_{3}$ respectively. The $A$ sublattice of the honeycomb lattice is formed by the red dots which in turn form the diamond lattice, whose bonds are denoted by the red dashed lines. The unit vectors of the diamond lattice is $\hat{a}_{1}$ and $\hat{a}_{2}$. After defining the Nambu spinor, the link variables form horizontal zig-zag chain. The sites in one of the chain can be marked by integer numbers $2k-1$, $2k$,..., with $A$ sublattice sites labelled by even numbers. Link variables on each zig-zag chain are mapped into spin variables defined on the $\hat{a}_{1}$ bonds of the diamond lattice, labelled by black dots. Spins corresponding to the same zig-zag chain form a horizontal line, which is the black dashed line.}
\label{fighoneycomblattice}
\end{figure}

\subsubsection{The model under SO(3) Majorana representation}

Following our definition in Sec. \ref{secso3}, we now turn to study the quantum XY model on honeycomb lattice (see Fig. \ref{fighoneycomblattice}) using the SO(3) Majorana representation. We introduce three types of Majorana fermions $\eta^{x},\eta^{y},\eta^{z}$ on each site to represent spins in the model. For each site, we pair up Majorana fermion $\eta^{x}$ and $\eta^{y}$ to form complex fermion $c$ according to Eq. (\ref{complexfermionxy}). Then, based on Eq. (\ref{XYinso3}), the Hamiltonian of quantum XY model on honeycomb lattice under the SO(3) Majorana representation is given by 
\begin{equation}
\label{HXYhoneycomb}
\mathcal{H}=J\sum_{\langle ij\rangle}(-\gamma_{i}\gamma_{j})(c_{i}^{\dagger}c_{j}+c_{i}c_{j}^{\dagger}),
\end{equation}
in which $i$ and $j$ are sites of the honeycomb lattice and $\langle ij\rangle$ denotes the bonds of the lattice. The three types of bonds of the honeycomb lattice are labelled by vectors $\hat{e}_{1},\hat{e}_{2}, \hat{e}_{3}$ and the two primitive vectors are denoted by $\hat{a}_{1}, \hat{a}_{2}$, as shown in Fig. \ref{fighoneycomblattice}. The $c_{i}$ fermions are formed by the Majorana fermion $\eta_{i}^{x}$ and $\eta_{i}^{y}$ according to (\ref{complexfermionxy}), such definition leaves the $\eta_{i}^{z}$ Majorana fermion unpaired at this stage. From now on and throughout this section, we use hatted symbol $\hat{i}$ to label the sites (and also the position vectors) of the honeycomb lattice belonging to the $A$ sublattice (the red dots in Fig. \ref{fighoneycomblattice}). As discussed in Sec. \ref{sec1djordanwigner}, to fix the Hilbert space of the Majorana fermion, we have to introduce $\frac{N}{2}$ constraints. Here we choose to pair up each $\hat{e}_{3}$ bond (vertical bond in Fig. \ref{fighoneycomblattice}) and require that 
\begin{equation}
\label{constrainthoneycomb}
\gamma_{\hat{i}}\gamma_{\hat{i}+\hat{e}_{3}}=-i,
\end{equation}
in which $\gamma_{\hat{i}}$ is the SO(3) singlet of the Majorana representation defined in (\ref{so3singlet}) and $\hat{i}$ belongs to the $A$ sublattice. With (\ref{constrainthoneycomb}) the Hamitonian (\ref{HXYhoneycomb}) is transformed into
\begin{eqnarray}
\label{HXYhoneycomb1}
\begin{aligned}
\mathcal{H}=J\sum_{\hat{i}\in \langle A\rangle}&(-\gamma_{\hat{i}}\gamma_{\hat{i}+\hat{e}_{3}})c_{\hat{i}}^{\dagger}c_{\hat{i}+\hat{e}_{3}}+(\eta_{\hat{i}}^{z}\eta_{\hat{i}+\hat{e}_{1}}^{z})c_{\hat{i}}^{\dagger}c_{\hat{i}+\hat{e}_{1}}+\\&(\eta_{\hat{i}}^{z}\eta_{\hat{i}+\hat{e}_{2}}^{z})c_{\hat{i}}^{\dagger}c_{\hat{i}+\hat{e}_{2}}+\text{h.c.}\\
=J\sum_{\hat{i}\in \langle A\rangle}&(\eta_{\hat{i}}^{z}\eta_{\hat{i}+\hat{e}_{1}}^{z})(c_{\hat{i}}^{\dagger}c_{\hat{i}+\hat{e}_{1}}+c_{\hat{i}}c_{\hat{i}+\hat{e}_{1}}^{\dagger})+\\&(\eta_{\hat{i}}^{z}\eta_{\hat{i}+\hat{e}_{2}}^{z})(c_{\hat{i}}^{\dagger}c_{\hat{i}+\hat{e}_{2}}+c_{\hat{i}}c_{\hat{i}+\hat{e}_{2}}^{\dagger})+\\&i(c_{\hat{i}}^{\dagger}c_{\hat{i}+\hat{e}_{3}}+c_{\hat{i}}c_{\hat{i}+\hat{e}_{3}}^{\dagger}),
\end{aligned}
\end{eqnarray}
in which we have used the alternative form of XY spin interaction given by Eq. (\ref{XYinetaz}) for $\hat{e}_{1}$ bonds and $\hat{e}_{2}$ bonds.

For the next step, to simplify notation, we can pair up the complex fermions $c_{\hat{i}}$ and $c_{\hat{i}+\hat{e}_{3}}$ located on the two ends of each $\hat{e}_{3}$ bonds in the honeycomb lattice into {\it Nambu spinor} 
\begin{equation}
\label{nambuspinorhoneycomb}
\psi_{\hat{i}}=\left(\begin{array}{c}
c_{\hat{i}+\hat{e}_{3}}\\c_{\hat{i}}
\end{array}\right), \qquad \psi_{\hat{i}}^{\dagger}=\left(\begin{array}{cc}
c_{\hat{i}+\hat{e}_{3}}^{\dagger} & c_{\hat{i}}^{\dagger}.
\end{array}\right)
\end{equation} 
The positions of the Nambu spinors are chosen to be the sites of the $A$ sublattice. Using the Nambu spinors we have that
\begin{eqnarray}
\begin{aligned}
&c_{\hat{i}}^{\dagger}c_{\hat{i}+\hat{e}_{3}}+c_{\hat{i}}c_{\hat{i}+\hat{e}_{3}}^{\dagger}=\psi_{\hat{i}}^{\dagger}\left(\begin{array}{cc}
0&-1\\ 1&0
\end{array}\right)\psi_{\hat{i}},\\
&c_{\hat{i}}^{\dagger}c_{\hat{i}+\hat{e}_{1}}+c_{\hat{i}}c_{\hat{i}+\hat{e}_{1}}^{\dagger}=\psi_{\hat{i}}^{\dagger}\left(\begin{array}{cc}
0&0\\ 1&0
\end{array}\right)\psi_{\hat{i}+\hat{a}_{1}}-\text{h.c.},\\
&c_{\hat{i}}^{\dagger}c_{\hat{i}+\hat{e}_{2}}+c_{\hat{i}}c_{\hat{i}+\hat{e}_{2}}^{\dagger}=\psi_{\hat{i}}^{\dagger}\left(\begin{array}{cc}
0&0\\ 1&0
\end{array}\right)\psi_{\hat{i}+\hat{a}_{2}}-\text{h.c.}.
\end{aligned}
\end{eqnarray}
Using these relations, the Hamiltonian (\ref{HXYhoneycomb1}) can be transformed as
\begin{eqnarray}
\begin{aligned}
\label{Hhoneycomb1}
\mathcal{H}=J\sum_{\hat{i}\in \langle A\rangle}\bigg[&\frac{1}{2}\psi_{\hat{i}}^{\dagger}\tilde{\sigma}_{y}\psi_{\hat{i}}+(\eta_{\hat{i}}^{z}\eta_{\hat{i}+\hat{e}_{1}}^{z})\psi_{\hat{i}}^{\dagger}\left(\begin{array}{cc}
0&0\\ 1&0
\end{array}\right)\psi_{\hat{i}+\hat{a}_{1}} \\&+(\eta_{\hat{i}}^{z}\eta_{\hat{i}+\hat{e}_{2}}^{z})\psi_{\hat{i}}^{\dagger}\left(\begin{array}{cc}
0&0\\ 1&0
\end{array}\right)\psi_{\hat{i}+\hat{a}_{2}}   \bigg]+\text{h.c.},
\end{aligned}
\end{eqnarray}
in which $\tilde{\sigma}_{y}$ is the Pauli matrix acting on the spin space of the Nambu spinor.
 
The link variables $\eta_{\hat{i}}^{z}\eta_{\hat{i}+\hat{e}_{1}}^{z}$ etc. in Eq. (\ref{Hhoneycomb1}) form a quasi-one-dimensional structure. In the honeycomb lattice, taking a horizontal zig-zag chain formed by $\hat{e}_{1}$ and $\hat{e}_{2}$, we see that there is a Majorana fermion $\eta^{z}$ on each site of the zig-zag chain (see Fig. \ref{fighoneycomblattice}). In the Hamiltonian (\ref{Hhoneycomb1}), Majorana fermions $\eta^{z}$ on different zig-zag chains do not talk to each other. For one specific horizontal zig-zag chain we label the sites in the following way: for site $\hat{i}$ on the A sublattice, we assign an even integer $2k$ to it; the site $\hat{i}+\hat{e}_{1}$ is assigned an odd integer $2k-1$ and the site $\hat{i}+\hat{e}_{2}$ the number $2k+1$, as shown in Fig. \ref{fighoneycomblattice}. The Majorana fermions $\eta^{z}$ on the zig-zag chain form a Kitaev chain \cite{kitaev2001}. Previously we paired up each $\hat{e}_{3}$ bonds to define the Nambu spinor in terms of the complex fermions formed by $\eta^{x}$ and $\eta^{y}$ Majorana fermions, we can pair up the independent $\eta^{z}$ Majorana fermions in a different way. Here, we choose to pair up the Majorana fermion $\eta^{z}$ on sites $2k-1$ and $2k$, in other words, sites $\hat{i}+\hat{e}_{1}$ and $\hat{i}$, and define complex fermion $d$, which we place on the middle point of the two paired sites, as 
\begin{equation}
d_{2k-\frac{1}{2}}=\frac{1}{2}(\eta_{2k-1}^{z}-i\eta_{2k}^{z}), \quad d_{2k-\frac{1}{2}}^{\dagger}=\frac{1}{2}(\eta_{2k-1}^{z}+i\eta_{2k}^{z}).
\end{equation}
Here, we temporarily use the assigned number to label sites in the horizontal zig-zag chain (see Fig. \ref{fighoneycomblattice}).

To make further progress, for the horizontal (zig-zag) chain, we can perform the 1d Jordan-Wigner transformation (see Sec. \ref{sec1djordanwigner}) for complex fermion $d_{2k-\frac{1}{2}}$ in the following way:
\begin{eqnarray}
\label{jordanwignerfordfermion}
\begin{aligned}
&d_{2k-\frac{1}{2}}=\sigma_{2k-\frac{1}{2}}^{-}e^{i\pi\sum_{j=1}^{k-1}\frac{1}{2}(1+\sigma_{2j-\frac{1}{2}}^{z})}, \\ &d_{2k-\frac{1}{2}}^{\dagger}=\sigma_{2k-\frac{1}{2}}^{+}e^{-i\pi\sum_{j=1}^{k-1}\frac{1}{2}(1+\sigma_{2j-\frac{1}{2}}^{z})},
\end{aligned}
\end{eqnarray}
with Jordan-Wigner spins defined on sites numbered $2k-\frac{1}{2}$, which is the mid-point of two integer-numbered sites: site $2k-1$ and site $2k$. Using these definition, the link variables in (\ref{Hhoneycomb1}), which in terms of $d$ fermion read $i\eta_{2k-1}^{z}\eta_{2k}^{z}=1-2d_{2k-\frac{1}{2}}^{\dagger}d_{2k-\frac{1}{2}}$ and $i\eta_{2k}^{z}\eta_{2k+1}^{z}=-(d_{2k-\frac{1}{2}}-d_{2k-\frac{1}{2}}^{\dagger})(d_{2k+\frac{3}{2}}+d_{2k+\frac{3}{2}}^{\dagger})$, can be transformed into
\begin{equation}
\label{honeycombmapping1}
i\eta_{2k-1}^{z}\eta_{2k}^{z} \rightarrow -\sigma^{z}_{2k-\frac{1}{2}}, \quad i\eta_{2k}^{z}\eta_{2k+1}^{z}\rightarrow \sigma_{2k-\frac{1}{2}}^{x}\sigma_{2k+\frac{3}{2}}^{x}.
\end{equation}
So far we have discussed only one chain, for other zig-zag chains we can pair up $\eta^{z}$ Majorana fermions in the same way and put the $d$ complex fermion and the Jordan-Wigner spins on the mid-points of all the $\hat{e}_{1}$ bonds. 

After the pairing of sites $\hat{i}$ and $\hat{i}+\hat{e}_{3}$ in our definition of Nambu spinor, the effective lattice for the Nambu spinors has become a {\it diamond shaped lattice} (or simply {\it diamond lattice}) whose sites are the $A$ sublattice points of the honeycomb lattice. In Fig. \ref{fighoneycomblattice}, the diamond lattice is formed by the red dots and we still use $\hat{i}$ to label the sites of the diamond lattice. The honeycomb bond $\langle \hat{i},\hat{i}+\hat{e}_{1}\rangle$ effectively becomes diamond bond $\langle \hat{i},\hat{i}+\hat{a}_{1}\rangle$. For the system of Nambu spinors, we can effectively put the Jordan-Wigner spins on the diamond lattice bonds $\langle \hat{i},\hat{i}+\hat{a}_{1}\rangle$. Using these notations the mapping (\ref{honeycombmapping1}) becomes 
\begin{equation}
\label{honeycombmapping}
i\eta_{\hat{i}+\hat{e}_{1}}^{z}\eta_{\hat{i}}^{z}\rightarrow -\sigma_{\hat{i}+\frac{1}{2}\hat{a}_{1}}^{z}, \quad i\eta_{\hat{i}}^{z}\eta_{\hat{i}+\hat{e}_{2}}^{z}\rightarrow \sigma_{\hat{i}+\frac{1}{2}\hat{a}_{1}}^{x}\sigma_{\hat{i}+\hat{a}_{2}-\frac{1}{2}\hat{a}_{1}}^{x}.
\end{equation}
On the other hand, according to QED in dimension $(2+1)$, we define the {\it conjugate Nambu spinor} $\bar{\psi}=\psi^{\dagger}\tilde{\sigma}_{y}$. Using (\ref{honeycombmapping}) and the definition of conjugate spinor, we transform the Hamiltonian (\ref{Hhoneycomb1}) into  
\begin{eqnarray}
\begin{aligned}
\label{Hhoneycomb2}
\mathcal{H}=&J\sum_{\hat{i}}\bigg[\frac{1}{2}\bar{\psi}_{\hat{i}}\psi_{\hat{i}}+\sigma_{\hat{i}+\frac{1}{2}\hat{a}_{1}}^{z}\bar{\psi}_{\hat{i}}\left(\begin{array}{cc}
-1&0\\ 0&0
\end{array}\right)\psi_{\hat{i}+\hat{a}_{1}}\\&+ \sigma_{\hat{i}+\frac{1}{2}\hat{a}_{1}}^{x}\sigma_{\hat{i}+\hat{a}_{2}-\frac{1}{2}\hat{a}_{1}}^{x}\bar{\psi}_{\hat{i}}\left(\begin{array}{cc}
-1&0\\ 0&0
\end{array}\right)\psi_{\hat{i}+\hat{a}_{2}}   \bigg]+\text{h.c.},
\end{aligned}
\end{eqnarray}
in which the summation is over every diamond lattice site $\hat{i}$.

\subsubsection{$Z_{2}$ gauge theory}

\begin{figure}
\includegraphics[width=0.4\textwidth]{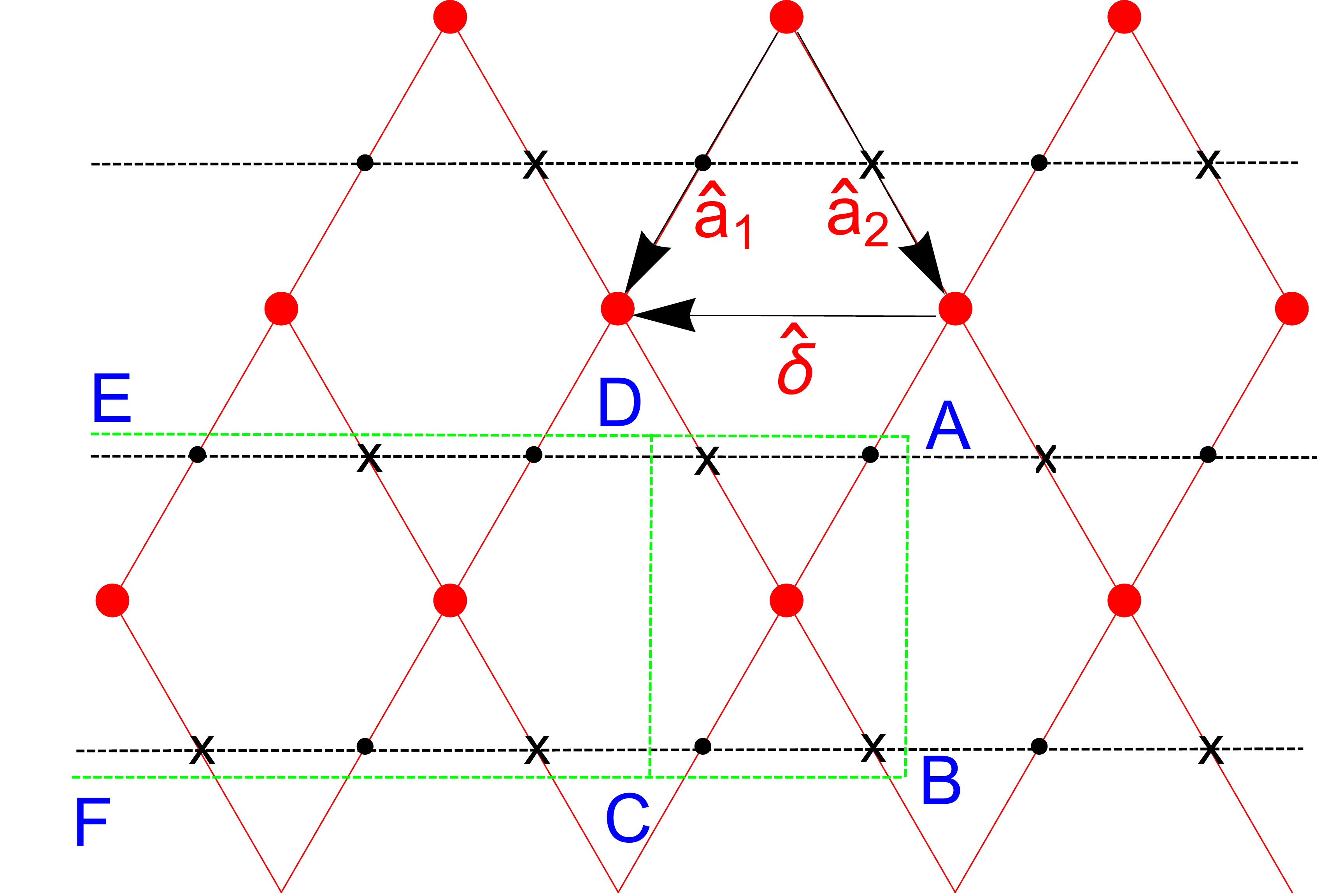}
\caption{The diamond lattice, with unit vector $\hat{a}_{1}$ and $\hat{a}_{2}$. Vector $\hat{\delta}$ is defined to be $\hat{a}_{1}-\hat{a}_{2}$. The spins from the link variables of the original honeycomb lattice are denoted by black dots in the $\hat{a}_{1}$ bonds, the original horizontal zig-zag chain in the honeycomb lattice become the horizontal black dashed lines. The treatment of the constraint (\ref{honeycombconstraint}) for site $\hat{i}$ in the middle of the (green dashed) block ABCD involves the spins in the half-infinite block CDEF. Duality transformation for each spin chain labelled by the black dashed line introduces new spin variables whose positions are denoted by black crosses. The Gauss law constraint for the $Z_{2}$ gauge theory involves the Nambu spinor and the four spin operators enclosed in ABCD.}
\label{figdiamondlattice}
\end{figure}

In order to fix the Hilbert space of the Majorana fermion, we have imposed the constraint (\ref{constrainthoneycomb}). Using the relation (\ref{tauandetaz}) and the definition of the Nambu spinor (\ref{nambuspinorhoneycomb}) we have the following relation
\begin{equation}
\gamma_{\hat{i}}\gamma_{\hat{i}+\hat{e}_{3}}=(-1)^{n_{\hat{i}}+n_{\hat{i}+\hat{e}_{3}}}\eta_{\hat{i}}^{z}\eta_{\hat{i}+\hat{e}_{3}}^{z}=(-1)^{\psi_{\hat{i}}^{\dagger}\psi_{\hat{i}}}\eta_{\hat{i}}^{z}\eta_{\hat{i}+\hat{e}_{3}}^{z},
\end{equation}
in which $n=c^{\dagger}c$ is the number of the complex fermion $c$. With this relation, the constraint can be rewritten as
\begin{equation}
\label{honeycombconstraint}
(-1)^{\psi_{\hat{i}}^{\dagger}\psi_{\hat{i}}}\eta_{\hat{i}}^{z}\eta_{\hat{i}+\hat{e}_{3}}^{z}=-i.
\end{equation}
In our previous discussion, we have taken the $\eta^{z}$ Majorana fermion on each horizontal zig-zag edge to form a Kitaev chain and pair them up within the chain to form complex fermion $d$. In terms of the $d$ fermion, the $\eta^{z}$ Majorana fermion can be written as
\begin{equation}
\label{relationetaandd}
\eta_{\hat{i}}^{z}=i(d_{\hat{i}+\frac{1}{2}\hat{a}_{1}}-d_{\hat{i}+\frac{1}{2}\hat{a}_{1}}^{\dagger}),\eta_{\hat{i}+\hat{e}_{3}}^{z}=(d_{\hat{i}-\frac{1}{2}\hat{a}_{1}}+d_{\hat{i}-\frac{1}{2}\hat{a}_{1}}^{\dagger}).
\end{equation}
We then performed 1d Jordan-Wigner transformation for the $d$ complex fermion to define the spin variables on the middle points of the $\hat{a}_{1}$ bonds of the diamond lattice. In the process, the spins (denoted by small black dots in Fig. \ref{fighoneycomblattice}) belonging to the same zig-zag edge form horizontal lines that cross the edges of the diamond lattice (the black dashed line in Fig \ref{fighoneycomblattice} and Fig. \ref{figdiamondlattice}). Based on the definition of the Jordan-Wigner transformation (\ref{jordanwignerfordfermion}) and (\ref{relationetaandd}) we have, defining vector $\hat{\delta}=\hat{a}_{1}-\hat{a}_{2}$, 
\begin{eqnarray}
\label{mappingetazandspin}
\begin{aligned}
&\eta_{\hat{i}}^{z}\rightarrow i(\sigma^{-}_{\hat{i}+\frac{1}{2}\hat{a}_{1}}-\sigma^{+}_{\hat{i}+\frac{1}{2}\hat{a}_{1}})e^{i\pi\sum_{j\geq 1}\frac{1}{2}(1+\sigma^{z}_{\hat{i}+\frac{1}{2}\hat{a}_{1}+j\hat{\delta}})}\\
&\eta_{\hat{i}+\hat{e}_{3}}^{z}\rightarrow (\sigma^{-}_{\hat{i}-\frac{1}{2}\hat{a}_{1}}+\sigma^{+}_{\hat{i}-\frac{1}{2}\hat{a}_{1}})e^{i\pi\sum_{j\geq 1}\frac{1}{2}(1+\sigma^{z}_{\hat{i}-\frac{1}{2}\hat{a}_{1}+j\hat{\delta}})},
\end{aligned}
\end{eqnarray}
here and hereafter we use $j$ to denote an integer variable. Therefore we have
\begin{equation}
\label{mappingetazspin}
\eta_{\hat{i}}^{z}\eta_{\hat{i}+\hat{e}_{3}}^{z}=\sigma_{\hat{i}+\frac{1}{2}\hat{a}_{1}}^{y}\sigma_{\hat{i}-\frac{1}{2}\hat{a}_{1}}^{x}e^{i\pi\sum_{j\geq 1}[1+\frac{1}{2}(\sigma^{z}_{\hat{i}+\frac{1}{2}\hat{a}_{1}+j\hat{\delta}}+\sigma^{z}_{\hat{i}-\frac{1}{2}\hat{a}_{1}+j\hat{\delta}})]}.
\end{equation}

To evaluate the phase factor in (\ref{mappingetazspin}), we note that in Fig. \ref{figdiamondlattice}, the $\sigma^{z}$ operators appearing in the exponent in (\ref{mappingetazspin}) are denoted as the black dots enclosed in the half-infinite region CDEF for the site $\hat{i}$ enclosed in the square ABCD.  To make further progress, we have to make some assumptions about the boundary conditions. Let us suppose that the number of sites on the horizontal lines from the site $\hat{i}+\frac{1}{2}\hat{a}_{1}$ and $\hat{i}-\frac{1}{2}\hat{a}_{1}$ to the boundary are equal, which means that the boundary is parallel to vector $\hat{a}_{1}$. Under such assumption, we have the total number of $\sigma^{z}$ operators enclosed in the region CDEF is an even number, which we call $2\tilde{N}$. Suppose that among these spin operators $m$ take the value $-1$ (which implies that $2\tilde{N}-m$ take $+1$), then the phase factor in (\ref{mappingetazspin}) is $(-1)^{2\tilde{N}-m}=(-1)^{m}$. This means that under the specific boundary condition we have 
\begin{eqnarray}
\begin{aligned}
\label{usefulrelation3}
&e^{i\pi\sum_{j\geq 1}[1+\frac{1}{2}(\sigma^{z}_{\hat{i}+\frac{1}{2}\hat{a}_{1}+j\hat{\delta}}+\sigma^{z}_{\hat{i}-\frac{1}{2}\hat{a}_{1}+j\hat{\delta}})]}\\=&\prod_{j\geq 1}\sigma^{z}_{\hat{i}-\frac{1}{2}\hat{a}_{1}+j\hat{\delta}}\sigma^{z}_{\hat{i}+\frac{1}{2}\hat{a}_{1}+j\hat{\delta}}.
\end{aligned}
\end{eqnarray}
Using (\ref{usefulrelation3}) we have that the constraint (\ref{honeycombconstraint}) is mapped into 
\begin{equation}
\label{honeycombconstraint1}
(-1)^{\psi_{\hat{i}}^{\dagger}\psi_{\hat{i}}}\sigma_{\hat{i}+\frac{1}{2}\hat{a}_{1}}^{z}\sigma_{\hat{i}+\frac{1}{2}\hat{a}_{1}}^{x}\sigma_{\hat{i}-\frac{1}{2}\hat{a}_{1}}^{x}\prod_{j\geq1}\sigma^{z}_{\hat{i}-\frac{1}{2}\hat{a}_{1}+j\hat{\delta}}\sigma^{z}_{\hat{i}+\frac{1}{2}\hat{a}_{1}+j\hat{\delta}}=1.
\end{equation}

To make further progress, we note that the Jordan-Wigner spins on the diamond lattice form horizontal spin chains, corresponding to the horizontal zig-zag edges of the original honeycomb lattice. In Fig. \ref{figdiamondlattice}, the spin chains are denoted by black dashed lines. For each horizontal spin chain in the diamond lattice, we can perform a duality transformation among spins defined on the sites and spins defined on the bonds \cite{Kogut1979,fradkinbook}. Specifically, for a horizontal spin chain on the diamond lattice formed by sites $\hat{i}+\frac{1}{2}\hat{a}_{1}+j\hat{\delta}$ where $j$ is an integer, we define a new set of spin variables $\tau$ on the mid-points of the two neighbouring sites of the original chain, formed by sites $\hat{i}+\frac{1}{2}\hat{a}_{2}+j\hat{\delta}$, in the following way
\begin{eqnarray}
\begin{aligned}
\label{dualitymapping1}
&\tau_{\hat{i}+\frac{1}{2}\hat{a}_{2}}^{z}=\sigma_{\hat{i}+\frac{1}{2}\hat{a}_{1}}^{x}\sigma_{\hat{i}+\frac{1}{2}\hat{a}_{1}-\hat{\delta}}^{x},\\ &\tau_{\hat{i}+\frac{1}{2}\hat{a}_{2}}^{x}=\prod_{j\geq 0}\sigma_{\hat{i}+\frac{1}{2}\hat{a}_{1}+j\hat{\delta}}^{z}.
\end{aligned}
\end{eqnarray}
We emphasize that the location of the new set of spin is on the $\hat{a}_{2}$ bonds of the diamond lattice, they are labelled as black crosses in Fig. \ref{figdiamondlattice}.

Under such duality mapping, the Hamiltonian (\ref{Hhoneycomb2}) and the constraint (\ref{honeycombconstraint1}) are both simplified significantly. The Hamiltonian becomes
\begin{widetext}
\begin{equation}
\label{Hhoneycomb3}
\mathcal{H}=J\sum_{\hat{i}}\bigg[\frac{1}{2}\bar{\psi}_{\hat{i}}\psi_{\hat{i}}+\sigma_{\hat{i}+\frac{1}{2}\hat{a}_{1}}^{z}\bar{\psi}_{\hat{i}}\left(\begin{array}{cc}
-1&0\\ 0&0
\end{array}\right)\psi_{\hat{i}+\hat{a}_{1}}+ \tau_{\hat{i}+\frac{1}{2}\hat{a}_{2}}^{z}\bar{\psi}_{\hat{i}}\left(\begin{array}{cc}
-1&0\\ 0&0
\end{array}\right)\psi_{\hat{i}+\hat{a}_{2}}   \bigg]+\text{h.c.}.
\end{equation} 
\end{widetext}
It takes the form of a standard lattice gauge theory \cite{Kogut1979,Prosko2017,fradkinbook} in which the Nambu spinor couples to $Z_{2}$ gauge field. The constraint (\ref{honeycombconstraint1}) becomes
\begin{equation}
\label{honeycombconstraint2}
(-1)^{\psi_{\hat{i}}^{\dagger}\psi_{\hat{i}}}\sigma_{\hat{i}+\frac{1}{2}\hat{a}_{1}}^{x}\sigma_{\hat{i}-\frac{1}{2}\hat{a}_{1}}^{x}\tau_{\hat{i}+\frac{1}{2}\hat{a}_{2}}^{x}\tau_{\hat{i}-\frac{1}{2}\hat{a}_{2}}^{x}=-1.
\end{equation}
It takes the form of a standard $Z_{2}$ Gauss law \cite{fradkinbook}. 

The gauge symmetry of the system does not take the usual form. Specifically we note that the Jordan-Wigner transformation and the duality transformation make the $Z_{2}$ gauge symmetry somewhat non-local. The transformation $\psi_{\hat{i}}\rightarrow -\psi_{\hat{i}}$ in the matter field must accompany the following change in the $\sigma$ gauge field: $\sigma^{x,y}_{\hat{i}+\frac{1}{2}\hat{a}_{1}-j\hat{\delta}}\rightarrow -\sigma^{x,y}_{\hat{i}+\frac{1}{2}\hat{a}_{1}-j\hat{\delta}},\quad \sigma_{\hat{i}+\frac{1}{2}\hat{a}_{1}}^{y,z}\rightarrow -\sigma_{\hat{i}+\frac{1}{2}\hat{a}_{1}}^{y,z}$ and $\sigma^{x,y}_{\hat{i}-\frac{1}{2}\hat{a}_{1}-j\hat{\delta}}\rightarrow -\sigma^{x,y}_{\hat{i}-\frac{1}{2}\hat{a}_{1}-j\hat{\delta}}.\quad \sigma_{\hat{i}-\frac{1}{2}\hat{a}_{1}}^{x,z}\rightarrow -\sigma_{\hat{i}-\frac{1}{2}\hat{a}_{1}}^{x,z}$, in which integer $j=1,2,3,...$; the transformation for $\tau$ spin can be deduced from Eq. (\ref{dualitymapping1}). Although the gauge transformation involves half-infinite spin chains, the only relevant change that manifests in the Hamiltonian (\ref{Hhoneycomb3}) is the following: $\psi_{\hat{i}}\rightarrow -\psi_{\hat{i}}$ and $\sigma^{z}_{\hat{i}\pm \frac{1}{2}\hat{a}_{1}}\rightarrow -\sigma^{z}_{\hat{i}\pm \frac{1}{2}\hat{a}_{1}},\quad \tau^{z}_{\hat{i}\pm \frac{1}{2}\hat{a}_{2}}\rightarrow -\tau^{z}_{\hat{i}\pm \frac{1}{2}\hat{a}_{2}}$, which is local. For all $\hat{i}$, the gauge transformation results in a sign change for even number of spins in the constraint (\ref{honeycombconstraint2}), thus leaves it invariant.

Despite the simple form of the $Z_{2}$ Hamitonian (\ref{Hhoneycomb3}) and the Gauss law constraint (\ref{honeycombconstraint2}), the model is still not solvable because the nontrivial relations between the $Z_{2}$ gauge field (\ref{dualitymapping1}), they are not independent from each other and thus we cannot fix the gauge in the usual way. 

Using the constraint (\ref{honeycombconstraint2}) we can define a projection operator for each site $\hat{i}$,
\begin{equation}
\label{projectorhoneycomb}
\mathcal{P}_{\hat{i}}=\frac{1}{2}[(-1)^{\psi_{\hat{i}}^{\dagger}\psi_{\hat{i}}}\sigma_{\hat{i}+\frac{1}{2}\hat{a}_{1}}^{x}\sigma_{\hat{i}-\frac{1}{2}\hat{a}_{1}}^{x}\tau_{\hat{i}+\frac{1}{2}\hat{a}_{2}}^{x}\tau_{\hat{i}-\frac{1}{2}\hat{a}_{2}}^{x}-1].
\end{equation}
It can be proved that the projector on each site commutes with the Hamiltonian (\ref{Hhoneycomb3}), $[\mathcal{P}_{\hat{i}},\mathcal{H}]=0$. This can be seen by noting that in the original definition of the constraint (\ref{constrainthoneycomb}) the operators $\gamma_{\hat{i}}\gamma_{\hat{i}+\hat{e}_{3}}$ commute with the original spin Hamiltonian. The Hamiltonian (\ref{Hhoneycomb3}) is defined in an enlarged Hilbert space. To get to the physical Hilbert space, we have to use the projection operator to project the state 
\begin{equation}
|\psi_{\text{phys}}\rangle=\prod_{\hat{i}}\mathcal{P}_{\hat{i}}|\psi\rangle,
\end{equation}
in which $|\psi\rangle$ is any state in the enlarged Hilbert space and the projected state $|\psi_{\text{phys}}\rangle$ is in the physical space. 

Because the projectors commute with the Hamiltonian, if we manage to find the eigenvalues of the Hamitonian (\ref{Hhoneycomb3}) in the enlarged Hilbert space, the true spectrum of the system will be the same. Unfortunately, as mentioned before, the spectrum of (\ref{Hhoneycomb3}) is hard to find even in the enlarged Hilbert space because of the non-trivial relation of the gauge fields (\ref{dualitymapping1}). The duality mapping (\ref{dualitymapping1}) does not allow us to simply pick up a gauge like $\sigma^{z}=1$ and $\tau^{z}=1$ for all the bonds, therefore exact solution of the spectrum is unavailable.

\subsection{The $90^{\circ}$ Compass Model on Square Lattice} \label{seccompass}

\begin{figure}
\includegraphics[width=0.4\textwidth]{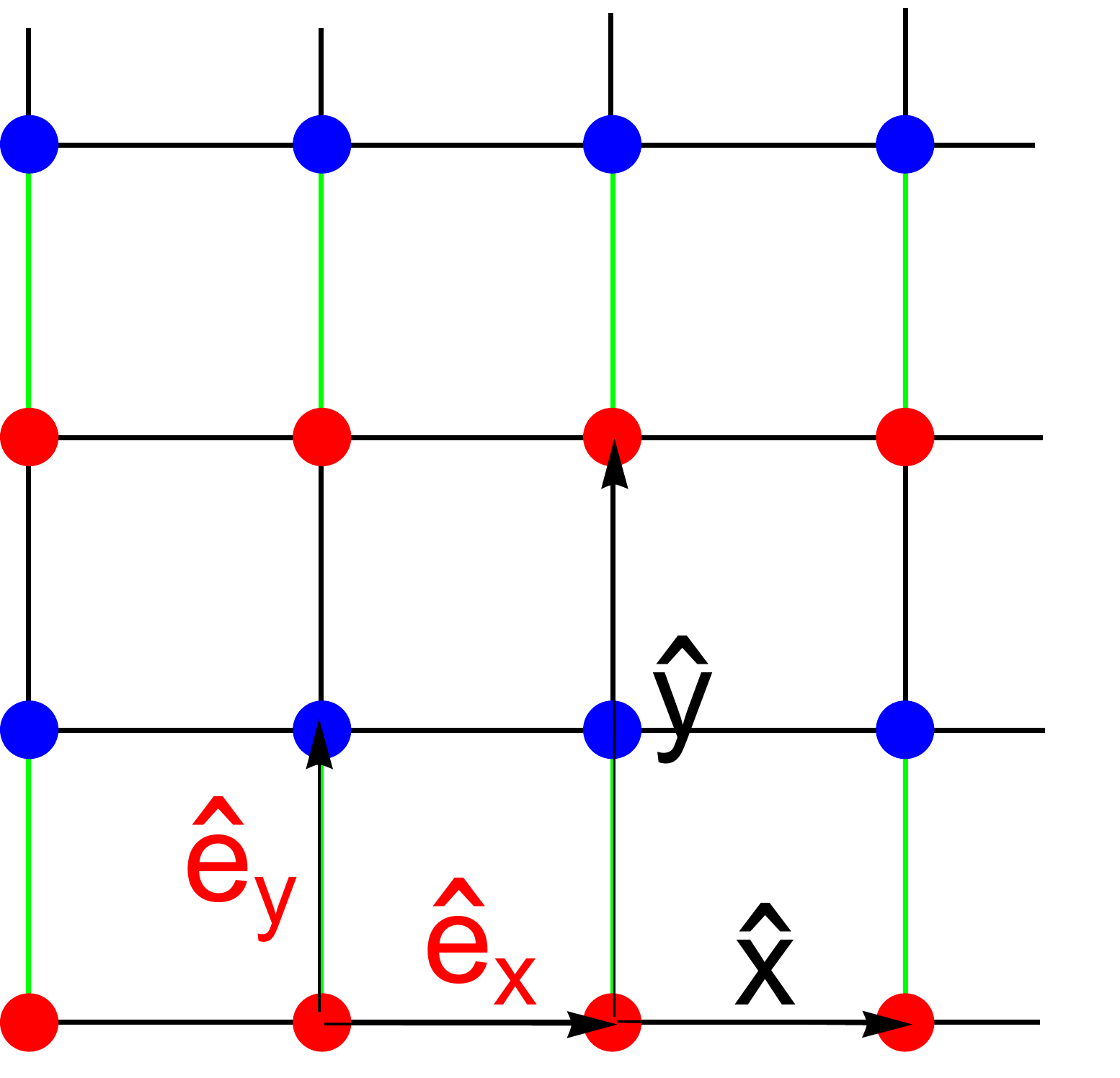}
\caption{The square lattice, with unit vectors $\hat{e}_{x}$ and $\hat{e}_{y}$. Spins in the $90^{\circ}$ compass are defined on the sites of the square lattice. Under SO(3) Majorana representation, we pair up the green bonds to form complex fermion. After the pairing, the lattice breaks into $A$ sublattice labelled by the red dots, and $B$ sublattice labelled by the blue dots. Complex fermion is defined on the $A$ sublattice, which then forms a rectangle lattice. The unit vectors of the rectangle lattice are labelled by $\hat{x}$ and $\hat{y}$.}
\label{figsquarelattice}
\end{figure}

\subsubsection{The model and SO(3) Majorana representation}

The {\it compass models} refer to a group of frustrated lattice spin models in which the spin interaction is bond-dependent (for a review, see Ref. \onlinecite{Nussinov2015}). On the two-dimensional square lattice, the bonds can be categorized by its direction, as shown in Fig. \ref{figsquarelattice}, we call the horizontal bonds in the lattice {\it x-bonds} and vertical bonds {\it y-bonds}. In the $90^{\circ}$ compass model on 2d square lattice \cite{Nussinov2015,chen2007,Nussinov2006}, the spins are placed on each site of the square lattice and only the x-components are interacting on x-bonds and only y-components are interacting on the y-bonds. Correspondingly, the Hamiltonian is given by 
\begin{equation}
\label{compasshamiltonian}
\mathcal{H}=\sum_{\langle ij\rangle_{x}}J_{1}\sigma_{i}^{x}\sigma_{j}^{x}+\sum_{\langle ij\rangle_{y}}J_{2}\sigma_{i}^{y}\sigma_{j}^{y},
\end{equation}
in which $\langle ij\rangle_{x}$ denotes the x-bonds, and $\langle ij\rangle_{y}$ denotes the y-bonds, and $J_{1}$ and $J_{2}$ are the coupling strength on x-bonds and y-bonds respectively.  

Following our discussion in Sec. \ref{secso3}, we can use the SO(3) Majorana representation to study this model. The first step is to use three Majorana fermions $\eta_{i}^{\alpha}$ with $\alpha=x,y,z$ to represent each spin operator. Using the definition of the SO(3) Majorana representation in Eq. (\ref{so3majoranarep}) we have
\begin{equation}
\label{squaredecomposition}
\sigma_{i}^{x}\sigma_{j}^{x}=(\eta_{i}^{y}\eta_{j}^{y})(\eta_{i}^{z}\eta_{j}^{z}),\quad
\sigma_{i}^{y}\sigma_{j}^{y}=(\eta_{i}^{z}\eta_{j}^{z})(\eta_{i}^{x}\eta_{j}^{x}).
\end{equation}
According to the Hamiltonian (\ref{compasshamiltonian}), such decomposition into Majorana fermions implies that the $\eta^{x}$ and $\eta^{y}$ Majorana fermions only hop on each y and x-bond respectively and the $\eta^{z}$ Majorana fermions hop on the entire lattice. Because the hopping of $\eta^{z}$ Majorana fermion on x and y bonds mutually commute, it is expected that dimensional reduction exists in this model \cite{Nussinov2015,chen2007,Nussinov2006}. 

For the next step we pair up the sites and define complex fermion operators. Here we choose to pair {\it half of the y-bonds}. In Fig. \ref{figsquarelattice}, the paired bonds are denoted by the green bonds. After the pairing, the lattice rotational symmetry is broken and the lattice contains two sublattices. The lower sites on the paired y-bonds are defined to be the $A$ sublattice and the upper sites are the $B$ sublattice. We then pair up the Majorana fermions on each paired bond to form three flavors of complex fermion,
\begin{equation}
\label{squarecomplexfermion}
c_{\hat{i}}^{\alpha}=\frac{1}{2}(\eta_{\hat{i}}^{\alpha}-i\eta_{\hat{i}+\hat{e}_{y}}^{\alpha}),\quad c_{\hat{i}}^{\alpha\dagger}=\frac{1}{2}(\eta_{\hat{i}}^{\alpha}+i\eta_{\hat{i}+\hat{e}_{y}}^{\alpha}),
\end{equation}
in which $\alpha=x,y,z$ and the position of these complex fermions is chosen to be on the $A$ sublattice. Here and hereafter, we use hatted symbol $\hat{i}$ to label sites of the $A$ sublattice of the original square lattice. Note that this definition of complex fermions is different from the one we used in Sec. \ref{secso3} and Sec. \ref{secxyhoneycomb}. With this definition of pairing and complex fermions, the lattice is effectively transformed into a {\it rectangle lattice} in which only the $A$ sublattice sites of the original square lattice are kept. The unit vectors of the original square lattice are labelled by $\hat{e}_{x}$ and $\hat{e}_{y}$ respectively. On the contrary, in the effective rectangle lattice, the unit vector of the y direction becomes $\hat{y}=2\hat{e}_{y}$ while the unit vector on the x direction is $\hat{x}=\hat{e}_{x}$ (see Fig. \ref{figsquarelattice}). We will use $\hat{x}$ and $\hat{y}$ to label the unit vectors as well as bonds on the rectangle lattice.

In order to fix the Hilbert space of the Majorana fermions, we require that for each paired bond $\gamma_{\hat{i}}\gamma_{\hat{i}+\hat{e}_{y}}=i$, with $\gamma_{\hat{i}}$ being the SO(3) singlet in the Majorana representation defined in (\ref{so3singlet}). In terms of the complex fermions, it reads
\begin{eqnarray}
\begin{aligned}
\label{condition}
\gamma_{\hat{i}}\gamma_{\hat{i}+\hat{e}_{y}}&=-i(2c_{\hat{i}}^{x\dagger}c_{\hat{i}}^{x}-1)(2c_{\hat{i}}^{y\dagger}c_{\hat{i}}^{y}-1)(2c_{\hat{i}}^{z\dagger}c_{\hat{i}}^{z}-1)\\&=i(-1)^{n_{\hat{i}}^{x}+n_{\hat{i}}^{y}+n_{\hat{i}}^{z}}=i,
\end{aligned}
\end{eqnarray}
in which we use $n_{\hat{i}}^{\alpha}=c_{\hat{i}}^{\alpha\dagger}c_{\hat{i}}^{\alpha}$ to denote the number of complex fermion of each flavor. The condition (\ref{condition}) implies that {\it there are even number of complex fermion on each site}.

Using the complex fermions (\ref{squarecomplexfermion}) and decomposition (\ref{squaredecomposition}) we can transform the Hamiltonian (\ref{compasshamiltonian}), which is first expressed as $\mathcal{H}=\mathcal{H}_{x}+\mathcal{H}_{y}$, in which $\mathcal{H}_{x}$ contains the spin interaction on $\hat{x}$-bonds and $\mathcal{H}_{y}$ contains spin interaction on $\hat{y}$-bonds. We have
\begin{eqnarray}
\label{Hx}
\begin{aligned}
\mathcal{H}_{x}=\sum_{\hat{i}\in \langle A\rangle}&J_{1}(\sigma_{\hat{i}}^{x}\sigma_{\hat{i}+\hat{e}_{x}}^{x}+\sigma_{\hat{i}+\hat{e}_{y}}^{x}\sigma_{\hat{i}+\hat{e}_{x}+\hat{e}_{y}}^{x})\\
=\sum_{\hat{i}}&2J_{1}[(c_{\hat{i}}^{y}c_{\hat{i}+\hat{x}}^{y}+c_{\hat{i}}^{y\dagger}c_{\hat{i}+\hat{x}}^{y\dagger})(c_{\hat{i}}^{z}c_{\hat{i}+\hat{x}}^{z}+c_{\hat{i}}^{z\dagger}c_{\hat{i}+\hat{x}}^{z\dagger})\\&+(c_{\hat{i}}^{y\dagger}c_{\hat{i}+\hat{x}}^{y}+c_{\hat{i}}^{y}c_{\hat{i}+\hat{x}}^{y\dagger})(c_{\hat{i}}^{z\dagger}c_{\hat{i}+\hat{x}}^{z}+c_{\hat{i}}^{z}c_{\hat{i}+\hat{x}}^{z\dagger})].
\end{aligned}
\end{eqnarray}
And 
\begin{eqnarray}
\label{Hy}
\begin{aligned}
\mathcal{H}_{y}=&\sum_{\hat{i}\in \langle A\rangle}J_{2}(\sigma_{\hat{i}}^{y}\sigma_{\hat{i}+\hat{e}_{y}}^{y}+\sigma_{\hat{i}+\hat{e}_{y}}^{y}\sigma_{\hat{i}+2\hat{e}_{y}}^{y})\\
=&\sum_{\hat{i}}J_{2}[(2c_{\hat{i}}^{y\dagger}c_{\hat{i}}^{y}-1)\\&-(c_{\hat{i}}^{x}-c_{\hat{i}}^{x\dagger})(c_{\hat{i}+\hat{y}}^{x}+c_{\hat{i}+\hat{y}}^{x\dagger})(c_{\hat{i}}^{z}-c_{\hat{i}}^{z\dagger})(c_{\hat{i}+\hat{y}}^{z}+c_{\hat{i}+\hat{y}}^{z\dagger})].
\end{aligned}
\end{eqnarray}
In the equations above, we have used the condition (\ref{condition}). In both (\ref{Hx}) and (\ref{Hy}), the second summation is done for the sites in the retangle lattice, which coincide with the A sublattice of the original square lattice and thus are also labelled by $\hat{i}$.

Notice from Eq. (\ref{Hx}) and Eq. (\ref{Hy}) that the complex fermions $c^{x}$ and $c^{y}$ only hop within each individual chain of y-bonds and x-bonds respectively. Fermions on different chains don't talk to each other, which implies that the dynamics of the complex fermions $c_{i}^{x}$ and $c_{i}^{y}$ is {\it quasi-one-dimensional}. This invites us to perform 1d Jordan-Wigner transformation for fermions $c_{i}^{x}$ and $c_{i}^{y}$.

\subsubsection{Jordan-Wigner transformation for complex fermions and duality transformation} 

\begin{figure}
\includegraphics[width=0.4\textwidth]{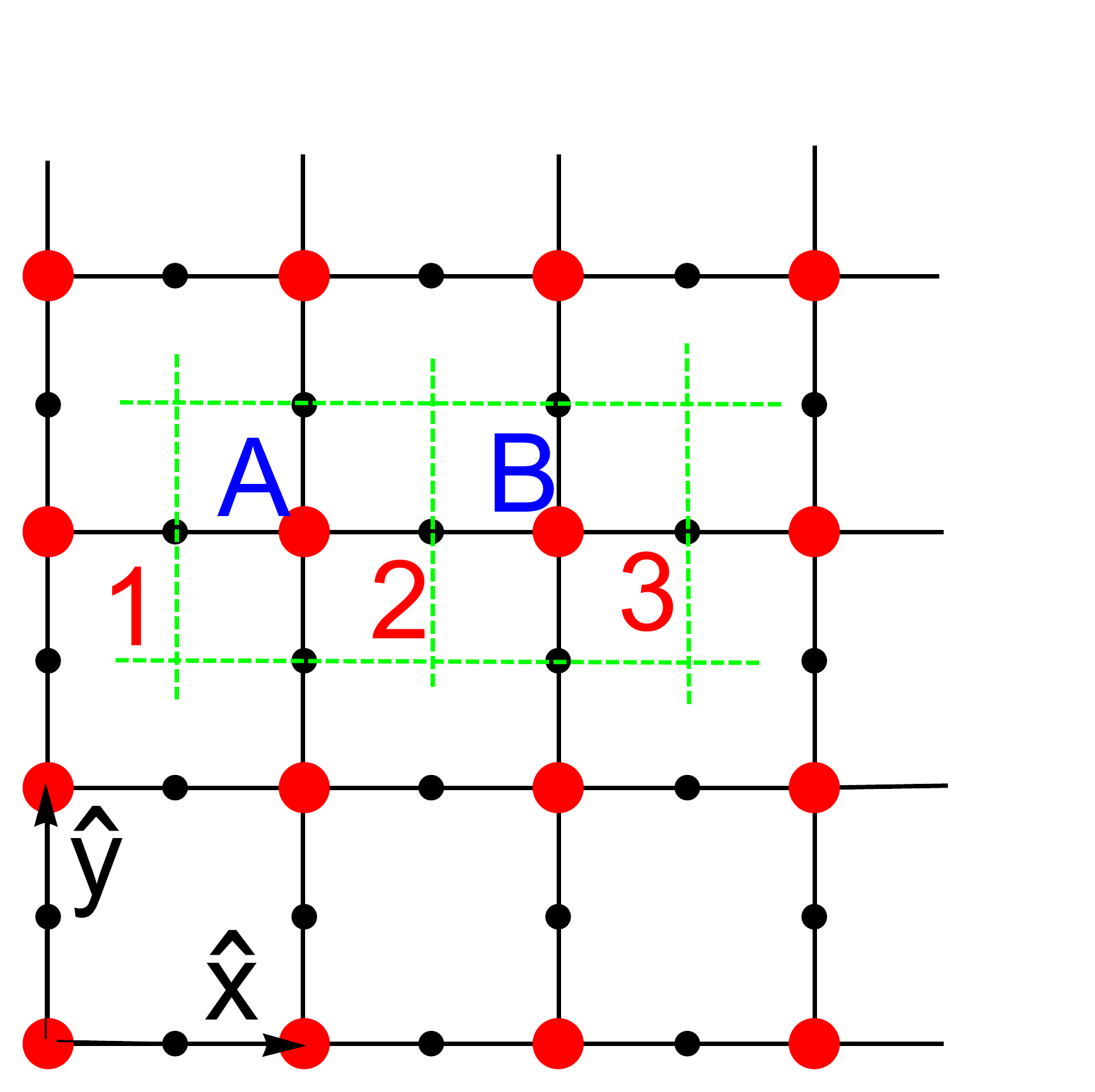}
\caption{The rectangle lattice, with unit vectors $\hat{x}$ and $\hat{y}$, note that we have shrinked the $\hat{y}$ to half of its length to achieve a clearer look. The complex fermion lives on the sites of the rectangle lattice, denoted by the red dots. The spin variables $\tau$ are defined on the bonds of the lattice, which are the black dots. In the $Z_{2}$ gauge theory Hamiltonian (\ref{totalH}), the hopping of complex fermions defined on points A and B couples to the spins on sites 1, 2, and 3. The dual lattice can be defined by connecting the centers of plaquettes of the original lattice. Part of the dual lattice is shown by the green dashed lines.}
\label{figrectanglelattice}
\end{figure}

According to the one-dimensional Jordan-Wigner transformation (\ref{jordanwigner1d}), for a 1d chain of complex fermions $c_{i}$ we can define a chain of spins $\sigma_{i}$. Specifically in our case, for each site $\hat{i}$ we have to define two sets of spin variables: one for the Jordan-Wigner transformation of $c_{\hat{i}}^{y}$ fermions on x-axis, which we call $\sigma_{\hat{i}}$; the other one for the Jordan-Wigner transformation of $c_{\hat{i}}^{x}$ fermions on y-axis, which we call $\tilde{\sigma}_{\hat{i}}$. (The $c^{z}$ fermion and the Jordan-Wigner spin variables $\sigma$ and $\tilde{\sigma}$ are located at sites of the rectangle lattice, labelled by red dots in Fig. \ref{figrectanglelattice}.) Using (\ref{jordanwigner1d}) we have the Jordan-Wigner transformation on $\hat{x}$-bonds,
\begin{eqnarray}
\label{xbosonization}
\begin{aligned}
& c_{\hat{i}}^{y}c_{\hat{i}+\hat{x}}^{y}+c_{\hat{i}}^{y\dagger}c_{\hat{i}+\hat{x}}^{y\dagger}\rightarrow \frac{i}{2}(\sigma_{\hat{i}}^{y}\sigma_{\hat{i}+\hat{x}}^{x}+\sigma_{\hat{i}}^{x}\sigma_{\hat{i}+\hat{x}}^{y})\\
& c_{\hat{i}}^{y\dagger}c_{\hat{i}+\hat{x}}^{y}+c_{\hat{i}}^{y}c_{\hat{i}+\hat{x}}^{y\dagger}\rightarrow \frac{i}{2}(\sigma_{\hat{i}}^{y}\sigma_{\hat{i}+\hat{x}}^{x}-\sigma_{\hat{i}}^{x}\sigma_{\hat{i}+\hat{x}}^{y}).
\end{aligned}
\end{eqnarray}  
The Jordan-Wigner transformation on $\hat{y}$-bonds reads,
\begin{equation}
\label{ybosonization}
(c_{\hat{i}}^{x}-c_{\hat{i}}^{x\dagger})(c_{\hat{i}+\hat{y}}^{x}+c_{\hat{i}+\hat{y}}^{x\dagger})\rightarrow -\tilde{\sigma}_{\hat{i}}^{x}\tilde{\sigma}_{\hat{i}+\hat{y}}^{x}.
\end{equation}

At this stage, it is important to consider the constraint (\ref{condition}) in the form of the Jordan-Wigner spin variables $\sigma$ and $\tilde{\sigma}$. We have, based on the Jordan-Wigner transformation (\ref{jordanwigner1d}), that the number of fermion on each site is transformed according to
\begin{equation}
n_{\hat{i}}=c_{\hat{i}}^{\dagger}c_{\hat{i}}\rightarrow \sigma^{+}_{\hat{i}}\sigma^{-}_{\hat{i}}=\frac{1}{2}(1+\sigma^{z}_{\hat{i}}).
\end{equation} 
Thus the condition that there are even number of fermion on each site (see Eq. (\ref{condition})) is transformed into the following condition in terms of the number of the $c^{z}$ fermion and the two Jordan-Wigner spins on each site, 
\begin{equation}
2n_{\hat{i}}^{z}+\sigma^{z}_{\hat{i}}+\tilde{\sigma}^{z}_{\hat{i}}=\pm 2,
\end{equation}
or in another form 
\begin{equation}
\label{condition1}
(-1)^{n_{\hat{i}}^{z}}\sigma_{\hat{i}}^{z}\tilde{\sigma}_{\hat{i}}^{z}=1.
\end{equation}
Now we have two types of spin variables on each site, to simplify the problem we can apply a duality transformation of 1d spin system \cite{Kogut1979, fradkinbook} to transform the two types of spins on sites to spins on bonds.

To define the duality transformation, we introduce a new set of spin variables $\tau$ on the $\hat{x}$ and $\hat{y}$ bonds of the rectangle lattice. The new spin variables on $\hat{x}$-bonds $\tau_{\hat{i}+\frac{\hat{x}}{2}}$ are used to represent the $\sigma_{\hat{i}}$ variables and the new spin variables on $\hat{y}$-bonds $\tau_{\hat{i}+\frac{\hat{y}}{2}}$ are used to represent $\tilde{\sigma}_{\hat{i}}$ variables. In Fig. \ref{figrectanglelattice}, the new spin variables $\tau$ are denoted by the small black dots on the bonds. Due to the distinction between the $\tau$ variables on x-bonds and y-bonds (in contrast to the $\sigma$ and $\tilde{\sigma}$ variables which are located at the same site), there is no confusion in this transformation although we are using the same symbol to label all the new spin variables (for both $\sigma$ and $\tilde{\sigma}$). The duality transformation \cite{Kogut1979,fradkinbook} can be defined subsequently; in particular we have that the $\sigma_{\hat{i}}$ variables on x-bonds are transformed as
\begin{equation}
\sigma_{\hat{i}}^{z}=\tau_{\hat{i}-\frac{\hat{x}}{2}}^{x}\tau_{\hat{i}+\frac{\hat{x}}{2}}^{x}, \qquad \sigma_{\hat{i}}^{x}=\prod_{j\geq 1}\tau_{\hat{i}+\frac{\hat{x}}{2}-j\hat{x}}^{z}, 
\end{equation}
in which we use $j$ to denote an integer variable. Therefore we have
\begin{eqnarray}
\label{xduality}
\begin{aligned}
&\sigma_{\hat{i}}^{y}\sigma_{\hat{i}+\hat{x}}^{x}=-i\sigma_{\hat{i}}^{z}\sigma_{\hat{i}}^{x}\sigma_{\hat{i}+\hat{x}}^{x}=-\tau_{\hat{i}-\frac{\hat{x}}{2}}^{x}\tau_{\hat{i}+\frac{\hat{x}}{2}}^{y},\\
&\sigma_{\hat{i}}^{x}\sigma_{\hat{i}+\hat{x}}^{y}=i\sigma_{\hat{i}}^{x}\sigma_{\hat{i}+\hat{x}}^{x}\sigma_{\hat{i}+\hat{x}}^{z}=-\tau_{\hat{i}+\frac{\hat{x}}{2}}^{y}\tau_{\hat{i}+\frac{3\hat{x}}{2}}^{x}.
\end{aligned}
\end{eqnarray}
Similarly, the Jordan-Wigner spins on y axis are transformed as
\begin{equation}
\label{yduality}
\tilde{\sigma}_{\hat{i}}^{x}\tilde{\sigma}_{\hat{i}+\hat{y}}^{x}=\tau_{\hat{i}+\frac{\hat{y}}{2}}^{z},\qquad \tau_{\hat{i}-\frac{\hat{y}}{2}}^{x}\tau_{\hat{i}+\frac{\hat{y}}{2}}^{x}=\tilde{\sigma}_{\hat{i}}^{z}.
\end{equation}

With this duality transformation (\ref{xduality}) and (\ref{yduality}), the condition (\ref{condition}) (and further (\ref{condition1})) is transformed as
\begin{equation}
\label{condition2}
(-1)^{n_{\hat{i}}}\tau_{\hat{i}-\frac{\hat{x}}{2}}^{x}\tau_{\hat{i}+\frac{\hat{x}}{2}}^{x}\tau_{\hat{i}-\frac{\hat{y}}{2}}^{x}\tau_{\hat{i}+\frac{\hat{y}}{2}}^{x}\equiv 1.
\end{equation}

\subsubsection{$Z_{2}$ gauge theory}

Using the bond spin operators, the two parts of the Hamiltonina (\ref{Hx}) and (\ref{Hy}) can be transformed. First, using Jordan-Wigner transformation on x-axis (\ref{xbosonization}) and duality transformation (\ref{xduality}) we have
\begin{eqnarray}
\label{Hxtransformed}
\begin{aligned}
\mathcal{H}_{x}
\rightarrow \sum_{\hat{i}}&2J_{1}[\frac{i}{2}\tau_{\hat{i}+\frac{\hat{x}}{2}}^{y}(-\tau_{\hat{i}-\frac{\hat{x}}{2}}^{x}-\tau_{\hat{i}+\frac{3\hat{x}}{2}}^{x})(c_{\hat{i}}^{z}c_{\hat{i}+\hat{x}}^{z}+c_{\hat{i}}^{z\dagger}c_{\hat{i}+\hat{x}}^{z\dagger})\\&+\frac{i}{2}\tau_{\hat{i}+\frac{\hat{x}}{2}}^{y}(-\tau_{\hat{i}-\frac{\hat{x}}{2}}^{x}+\tau_{\hat{i}+\frac{3\hat{x}}{2}}^{x})(c_{\hat{i}}^{z\dagger}c_{\hat{i}+\hat{x}}^{z}+c_{\hat{i}}^{z}c_{\hat{i}+\hat{x}}^{z\dagger})]\\
=\sum_{\hat{i}}&J_{1}[(\tau_{\hat{i}+\frac{\hat{x}}{2}}^{x}\tau_{\hat{i}-\frac{\hat{x}}{2}}^{x})\tau_{\hat{i}+\frac{\hat{x}}{2}}^{z}(c_{\hat{i}}^{z}+c_{\hat{i}}^{z\dagger})(c_{\hat{i}+\hat{x}}^{z}+c_{\hat{i}+\hat{x}}^{z\dagger})\\&+(\tau_{\hat{i}+\frac{\hat{x}}{2}}^{x}\tau_{\hat{i}+\frac{3\hat{x}}{2}}^{x})\tau_{\hat{i}+\frac{\hat{x}}{2}}^{z}(c_{\hat{i}}^{z}-c_{\hat{i}}^{z\dagger})(c_{\hat{i}+\hat{x}}^{z}-c_{\hat{i}+\hat{x}}^{z\dagger})].
\end{aligned}
\end{eqnarray}
In the last equation we have used the fact that $aA+bB=\frac{1}{2}[(a+b)(A+B)+(a-b)(A-B)]$ for any variables $a,b$ and $A,B$. On the other hand, using Jordan-Wigner transformation (\ref{ybosonization}) and duality transformation (\ref{yduality}) we can transfrom Eq. (\ref{Hy}) as follows
\begin{equation}
\label{Hytransformed}
\mathcal{H}_{y}
\rightarrow \sum_{\hat{i}}J_{2}[\tau_{\hat{i}-\frac{\hat{x}}{2}}^{x}\tau_{\hat{i}+\frac{\hat{x}}{2}}^{x}+\tau_{\hat{i}+\frac{\hat{y}}{2}}^{z}(c_{\hat{i}}^{z}-c_{\hat{i}}^{z\dagger})(c_{\hat{i}+\hat{y}}^{z}+c_{\hat{i}+\hat{y}}^{z\dagger})],
\end{equation}
in which we have used the following transformation coming from the Jordan-Wigner transformation and duality transformation mentioned above: $(2c_{i}^{y\dagger}c_{i}^{y}-1)\rightarrow \sigma_{i}^{z}\rightarrow \tau_{i-\frac{\hat{x}}{2}}^{x}\tau_{i+\frac{\hat{x}}{2}}^{x}$.

Combining (\ref{Hxtransformed}) and (\ref{Hytransformed}) we can see that now the Hamiltonian involves complex fermion $c_{\hat{i}}^{z}$ defined on the sites of the rectangle lattice and spin variables $\tau^{\alpha}$ defined on the bonds of the rectangle lattice (see Fig. \ref{figrectanglelattice}). Part of the unphysical degrees of freedom are fixed by the gauge condition (\ref{condition2}) which takes the form of standard $Z_{2}$ Guass' law \cite{fradkinbook,fu2018}. Although the form of the transformed Hamiltonian is simple, it is not the usual $Z_{2}$ gauge theory \cite{Prosko2017} in that the bond variables contain non-commuting $\tau^{x}$ and $\tau^{z}$. 

For the next step, we can safely drop the index $z$ of the complex fermions without causing confusion since it is the only fermionic degree of freedom left. The total Hamiltonian is given by
\begin{eqnarray}
\label{totalH}
\begin{aligned}
\mathcal{H}=\sum_{\hat{i}}& J_{1}[(\tau_{\hat{i}+\frac{\hat{x}}{2}}^{x}\tau_{\hat{i}-\frac{\hat{x}}{2}}^{x})\tau_{\hat{i}+\frac{\hat{x}}{2}}^{z}(c_{\hat{i}}+c_{\hat{i}}^{\dagger})(c_{\hat{i}+\hat{x}}+c_{\hat{i}+\hat{x}}^{\dagger})\\&+(\tau_{\hat{i}+\frac{\hat{x}}{2}}^{x}\tau_{\hat{i}+\frac{3\hat{x}}{2}}^{x})\tau_{\hat{i}+\frac{\hat{x}}{2}}^{z}(c_{\hat{i}}-c_{\hat{i}}^{\dagger})(c_{\hat{i}+\hat{x}}-c_{\hat{i}+\hat{x}}^{\dagger})]\\&+J_{2}\tau_{\hat{i}-\frac{\hat{x}}{2}}^{x}\tau_{\hat{i}+\frac{\hat{x}}{2}}^{x}+J_{2}\tau_{\hat{i}+\frac{\hat{y}}{2}}^{z}(c_{\hat{i}}-c_{\hat{i}}^{\dagger})(c_{\hat{i}+\hat{y}}+c_{\hat{i}+\hat{y}}^{\dagger}).
\end{aligned}
\end{eqnarray}

To see the $Z_{2}$ gauge symmetry, we note that the Hamiltonian (\ref{totalH}) and the constraint (\ref{condition2}) are invariant under the transformation: $c_{\hat{i}}\rightarrow -c_{\hat{i}}$ and $\tau^{z}_{\hat{i}\pm \frac{\hat{x}}{2}}\rightarrow -\tau^{z}_{\hat{i}\pm \frac{\hat{x}}{2}},\quad \tau^{z}_{\hat{i}\pm \frac{\hat{y}}{2}}\rightarrow -\tau^{z}_{\hat{i}\pm \frac{\hat{y}}{2}}$, with all $\tau^{x}$ components unchanged. 

From the condition (\ref{condition2}) we can define a projector for each site $\hat{i}$,
\begin{equation}
\label{projectorsquare}
\mathcal{P}_{\hat{i}}=\frac{1}{2}[(-1)^{n_{\hat{i}}}\tau_{\hat{i}-\frac{\hat{x}}{2}}^{x}\tau_{\hat{i}+\frac{\hat{x}}{2}}^{x}\tau_{\hat{i}-\frac{\hat{y}}{2}}^{x}\tau_{\hat{i}+\frac{\hat{y}}{2}}^{x}+1].
\end{equation}
The projector (\ref{projectorsquare}) commutes with the Hamiltonian (\ref{totalH}), $[\mathcal{P}_{\hat{i}},\mathcal{H}]=0$ following the fact that $\{(-1)^{n_{i}},c_{i}\}=0$ and $\{(-1)^{n_{i}},c_{i}^{\dagger}\}=0$. It can also be seen by noting that in the first step the operators $\gamma_{\hat{i}}\gamma_{\hat{i}+\hat{e}_{y}}$ that are picked to define the condition (\ref{condition}) commute with the original spin Hamiltonian. The $Z_{2}$ Hamiltonian (\ref{totalH}) is defined in an enlarged Hilbert space, which contains the physical space as a subspace. The physical space is obtained by projection $|\psi_{\text{phys}}\rangle=\prod_{\hat{i}}\mathcal{P}_{\hat{i}}|\psi\rangle$, in which $|\psi\rangle$ is any state of the enlarged Hilbert space. Because we have $[\mathcal{P}_{\hat{i}},\mathcal{H}]=0$, if we manage to find the eigenstate of $\mathcal{H}$, the physical state will have the same energy after projection. This allows us to focus on the Hamiltonian (\ref{totalH}) first, find its eigenstates and its eigenvalues gives the exact energy spectrum of the model.

Unfortunately the $Z_{2}$ Hamiltonian is highly non-trivial. In Fig. \ref{figrectanglelattice}, we note that the hopping of complex fermions between sites A and B couples to $Z_{2}$ gauge fields on bonds 1, 2 and 3. In analogy to the U(1) lattice gauge theory \cite{fradkinbook}, the $\tau^{x}$ operator acts like electric field while the $\tau^{z}$ operator acts like magnetic vector potential. The non-trivial form of Hamiltonian (\ref{totalH}) means that the charge current in this $Z_{2}$ gauge theory couples non-trivially to the electric field. Another way to study the Hamiltonian is by going to the dual lattice which is defined by connecting the centers of all the plaquettes of the retangle lattice (in Fig. \ref{figrectanglelattice}, part of the dual lattice is shown by green dashed lines). On the dual lattice, we perform the duality transformation of electrical and magnetric fields, i.e. define a new set of fields $\tilde{\tau}^{z}=\tau^{x}$ and $\tilde{\tau}^{x}=-\tau^{z}$ on the same sites of the original fields. The new set of gauge fields $\tilde{\tau}$ are still defined on the bonds of the dual lattice; however, the charges, which become the magnetic monopoles after the transformation, are located at the centers of the plaquettes of the dual lattice and the condition (\ref{condition2}) becomes the flux attachment constraint to the magnetic monopole. 

\subsection{Discussion on the $Z_{2}$ gauge theories and the application of SO(3) Majorana representation in spin models} \label{secdiscussiononapplication}

To apply the SO(3) Majorana representation, we study two spin models on 2d lattices, namely the quantum XY model on honeycomb lattice and the $90^{\circ}$ compass model on square lattice. In both cases, we show how to use the SO(3) Majorana representation to exactly map the models into $Z_{2}$ lattice gauge theories. Specifically, we introduce $\frac{N}{2}$ constraints by pairing up sites and requiring that for each pair $\langle ij\rangle$, the value of the product of SO(3) singlet $\gamma_{i}\gamma_{j}$ is fixed. Due to the fact that these product operators commute with the spin Hamiltonian, we show that the conditions can be mapped into the standard form of Gauss law in the $Z_{2}$ gauge theories. Unfortunately, neither of the models is exactly solvable and the resulting $Z_{2}$ gauge theories are non-trivial in that we cannot simply pick up a gauge and determine the spectrum of the matter fields. To this end, further approximations are needed to treat these non-trivial $Z_{2}$ gauge theories. Here we give a brief discussion on the possible approximations that may be applied and make some remarks on future direction of study.

In the quantum XY model on honeycomb lattice, we obtain the exact $Z_{2}$ Hamiltonian (\ref{Hhoneycomb3}) with Gauss law constraint (\ref{honeycombconstraint2}). If we ignore the non-trivial relation between gauge field (\ref{dualitymapping1}) and set $\sigma^{z}=1$ and $\tau^{z}=1$ for all the bonds, we can get an approximate spectrum of the fermion. For that, we have to return to the language of complex fermion $c$. Due to the form of the Hamiltonian (\ref{Hhoneycomb3}) and the underlying lattice, the resulting spectrum is similar to graphene \cite{neto2009}. Adding a magnetic field to the model will corresponding to adding a chemical potential term to the complex fermion \cite{kumar2014}. On the other hand, using the approximate spectrum we can study the possible phase transition in the quantum XY model with finite temperature \cite{ding1990}. 

In the $90^{\circ}$ compass model on square lattice, the $Z_{2}$ Hamiltonian (\ref{totalH}) and the condition (\ref{condition2}) are exact results. To go further, we note that the link variables in (\ref{totalH}) on the $\hat{x}$ direction are still anti-commuting to each other. Such property is rooted in the anti-commuting link variables in the original Hamiltonian under SO(3) Majorana representation which was discussed in Sec. \ref{sec2djordanwigner}. We can apply mean-field theory to treat them and it is believed that proper mean-field treatment of (\ref{totalH}) will lead to comparable results as the previous works on this model \cite{Nussinov2015}, such as quantum phase transition near the point $J_{1}=J_{2}$ \cite{chen2007}. 

Our application of SO(3) Majorana representation in the two models in this paper and in the Kitaev model in previous study \cite{fu2018} shows a new way to treat spin models. This method features a series of exact mapping and the results are always $Z_{2}$ gauge theories with standard Gauss law. The exact $Z_{2}$ gauge theories contain all the physics of the original spin model and serve as the starting point of further approximations, if needed. At this stage, it is important to point out the limitation on the applicability of this method on spin models. As we seen in Sec. \ref{secso3}, in the SO(3) Majorana representation the z-component spin interaction is mapped into a four-fermion interaction (or density-density interaction), as shown in Eq. (\ref{JzpartofXXZ}). There is considerable difficulty in treating such four-fermion interaction \cite{Lee2006,shankar1994}. Therefore, the mapping of spin models to exact $Z_{2}$ gauge theories is only applicable to the spin Hamiltonians which do not have the spin rotational symmetry. Otherwise the four-fermion interaction is included and the application of SO(3) Majorana representation holds no advantage over other representations. Specifically, there is no ``$\sigma^{z}-\sigma^{z}$" interaction in either of the models considered here, and in the Kitaev model only one spin component is interacting on each bond \cite{Kitaev2006}. However, the exact condition on the applicability of the method is still lacking and one should consider the application of SO(3) Majorana representation in each individual spin model separately.   

\section{Conclusion and Outlook} \label{secconclusion}

In this work, we explore the properties of the SO(3) Majorana representation and discuss its application in two spin models. Being a non-local representation of spin, the SO(3) Majorana representation is compared with the Jordan-Wigner transformation in 1d and 2d. For a 1d spin chain, we find the SO(3) Majorana representation of spin corresponds to the Jordan-Wigner transformation under some specific conditions to fix the extra degrees of freedom in the Majorana Hilbert space. From that, we find that there is always some $Z_{2}$ redundancy in the application of the SO(3) Majorana representation if only $\frac{N}{2}$ fixing conditions are imposed on the Majorana Hilbert space (N is the number of spin in the system). We confirm this point in the studies of the two spin models where both models are exactly mapped into (non-trivial) $Z_{2}$ lattice gauge theories. Based on a lattice version of the Chern-Simons gauge theory, we find an equivalence between the SO(3) Majorana representation and the 2d Jordan-Wigner transformation (also known as the Chern-Simons JW transformation), such equivalence is not exact due to the limitation of the Chern-Simons JW transformation. Despite this, we are able to map the link variables in the SO(3) Majorana representation to the Wilson links in the Chern-Simons JW transformation (Eq. \ref{so3andcs}) provided that the lattice Chern-Simons gauge field is compactified with a Berry phase $e^{i\pi}$ every time $2\pi$ is added to the gauge field on each bond of the lattice. Moreover, we emphasize that the anticommuting link variables (link variables anticommute with each other whenever they share a vertex) are generally hard to handle in spin models. One can completely get rid of the anticommuting link variables only in some special cases (like the 1d spin chain and the Kitaev model \cite{fu2018}). In general, such anticommuting link variables either exist after we map the models into $Z_{2}$ gauge theories or result in some non-trivial feathers in the resulting theories. In order to treat these, some approximation is always needed. 

There are a few questions left unanswered in this work. First, as we noted before, the SO(3) Majorana representation can be applied to a broader range of models than the Jordan-Wigner transformation. Specifically in two-dimensional models, there are only limited cases where the Chern-Simons gauge theory can be defined on a lattice. To understand the origin of such limitation, further exploration of the lattice Chern-Simons gauge theory is needed. On the other hand, the SO(3) Majorana representation corresponds to the compactified $\text{U(1)}_{1}$ Chern-Simons Jordan-Wigner transformation with Berry phase $-1$, further studies are needed to explore the physical meaning of such Berry phase. Moreover, it is unclear if there is consistency between the quantized gauge field as a result of the compactification and the definition of the Berry phase in which the gauge field is made to ``go around" the closed manifold continuously and adiabatically. To this end, it is possible that including the Maxwell action in the lattice gauge theory would solve these questions. Furthermore, the SO(3) Majorana representation can be applied to any spatial dimension. In the three-dimensional space, there is some work on the Bose-Fermi transformation \cite{huerta1993}, further study is needed to clarify their relationship with the SO(3) Majorana representation. On the other hand, we should mention that there are other forms of the lattice Chern-Simons gauge theory \cite{Doucot2005a,Doucot2005b}. Their relation with the version we adopt here \cite{sun2015} still needs some clarification.

For the two spin models we considered in this work, we give little discussion on the further approximation needed to treat the $Z_{2}$ gauge theories and their implications on the physical properties of the models. Further exploration in this direction is left for future study. It is believed that, although the $Z_{2}$ gauge theories presented are not necessarily exactly solvable, the discrete nature of the gauge group will bring opportunities for us to have a better controlled way to study spin models (in contrast with the continuous gauge theories (or even non-abelian gauge theories) resulted from the slave-particle approach \cite{Lee2006,wen1991}). On the other hand, application of the SO(3) Majorana representation on other types of spin models, such as other types of compass models \cite{Nussinov2015}, is left for future study.

\section*{Acknowledgements}

The author thanks M. Voloshin for discussions on Chern-Simons gauge theory and N. Perkins for reading the manuscript. This work is supported by NSF DMR-1511768 Grant.

\appendix

\section{Hard-core Boson Representation of Spin}\label{appendixhardcore}

Spin can be viewed as hard-core boson which behave like bosonic operator but under the constraint that the number of boson on each site can only be 0 or 1 \cite{fradkinbook}. Here we start with a system of hard-core boson and study its properties. 

Suppose we have for each site $i$ a hard-core boson $a_{i}$. For any ordinary bosonic operator $b_{i}$, we have the commutation relation $[b_{i},b_{j}^{\dagger}]=\delta_{ij}$. However, this is not the commutation relation for hard-core boson $a_{i}$, for which we have to require that on each site there can only be 0 or 1 boson, in other words,
\begin{equation}
\label{hardcore1}
a_{i}^{2}=0; \qquad a_{i}^{\dagger 2}=0. 
\end{equation}
The Hilbert space for each hard-core boson is restricted to be spanned by two basis states $|0\rangle$ and $|1\rangle$. For an ordinary bosonic operator $b_{i}$, to go to this two-dimensional subspace of the original Hilbert space (which has inifinite dimension), a projection is needed. Let's call it $\hat{P}_{i}$. We have that the hard-core boson operator $a_{i}$ is obtained from the ordinary bosonic operator $b_{i}$ by $a_{i}=\hat{P}_{i}b_{i}\hat{P}_{i}$; $a_{i}^{\dagger}=\hat{P}_{i}b_{i}^{\dagger}\hat{P}_{i}$. Due to the fact that $[\hat{P}_{i},b_{i}]\neq 0$, the hard-core boson can be seen as {\it dressed boson}.

The hard-core boson has the following commutation relations
\begin{eqnarray}
\begin{aligned}
\label{commutationhardcoreboson}
&\{a_{i},a_{i}^{\dagger}\}=1; \qquad \text{and}  \\ &[a_{i},a_{j}]=[a_{i},a_{j}^{\dagger}]=[a_{i}^{\dagger},a_{j}^{\dagger}]=0 \qquad \text{for $i\neq j$}.
\end{aligned}
\end{eqnarray}
There is a one-to-one mapping between the Hilbert space of hard-core boson and the spin space, using the commutation relations of the hard-core boson operator (\ref{commutationhardcoreboson}), we have the following mapping between spin operator and hard-core boson operator,
\begin{eqnarray}
\begin{aligned}
\label{hardcorebosonspin}
&\sigma_{i}^{+}=\frac{1}{2}(\sigma_{i}^{x}+i\sigma_{i}^{y})=a_{i}^{\dagger}, \\ &\sigma_{i}^{-}=\frac{1}{2}(\sigma_{i}^{x}-i\sigma_{i}^{y})=a_{i}; \\ &\sigma_{i}^{z}=2a_{i}^{\dagger}a_{i}-1.
\end{aligned}
\end{eqnarray}

\section{Jordan-Wigner Transformation in 2d Using Chern-Simons Flux Attachment}\label{appendix2djordanwigner}

Here we follow Ref. \onlinecite{fradkin1989,fradkinbook,lopez1994,kumar2014} to give a brief review of the 2d Jordan-Wigner transformation using the Chern-Simons gauge theory. 
We start with a simple two-dimensional quantum XY model, which according to (\ref{hardcorebosonspin}) can be written in terms of hard-core boson as
\begin{equation}
\label{hardcorebosonh}
\mathcal{H}_{a}=\sum_{ij}a_{i}^{\dagger}a_{j}+\text{h.c.}.
\end{equation}
Here we assume the coupling constant $J=1$. Due to the exotic commutation relation of the hard-core boson (\ref{commutationhardcoreboson}), we will treat it as an anyonic operator.

On the other hand, we start with a fermionic system coupled to Chern-Simons gauge field. The fermions reside on the sites of the lattice while the gauge field is defined on the bonds or edges of the lattice. (If and only if the lattice has a one-to-one correspondence between sites and faces, the lattice CS gauge theory is well defined \cite{sun2015}.) The Hamiltonian is, setting coupling constant to unity, 
\begin{equation}
\label{Hxyincs}
\mathcal{H}_{f}=\sum_{ij}c_{i}^{\dagger}e^{iA_{ij}}c_{j}+\text{h.c.}.
\end{equation} 
The gauge field $A_{ij}$ is subject to Chern-Simons action (\ref{latticecs}), which results in a constraint (\ref{fluxattachonlattice}). On a certain lattice, such constraint can help us to solve the configuation of the gauge field classically according to the charge distribution of fermion $c$ \cite{fradkin1989,fradkinbook,wang1991,kumar2014}. To this end, if we define an operator 
\begin{equation}
\label{anyonandfermion}
\tilde{a}_{i}=e^{-i\phi_{i}}c_{i},
\end{equation}
in which operator $\phi_{i}$ is a functional of the density of the fermion $n_{i}=c_{i}^{\dagger}c_{i}$. Such functional form will lead to nontrivial commutation relation between operator $e^{-i\phi_{i}}$ and operator $c_{i}$. This will result in the exotic commutation relation of the anyonic operator $\tilde{a}_{i}$
\begin{equation}
\tilde{a}_{i}\tilde{a}_{j}^{\dagger}=\delta_{ij}-e^{i\delta}\tilde{a}_{j}^{\dagger}\tilde{a}_{i},
\end{equation} 
in which $\delta=\frac{\pi}{k}$ is a constant, with $k$ being the level of the Chern-Simons theory in (\ref{latticecs}) and (\ref{chernsimonsaction}). If the level $k=1$, then $\delta=\pi$ and the commutation relation of anyon $\tilde{a}$ becomes bosonic. Further more, it satisfies the hard-core condition (\ref{hardcore1}) following from its definition (\ref{anyonandfermion}). Therefore when $k=1$, the anyonic operator $\tilde{a}$ is identified to be a hard-core boson. 

Under the condition that $k=1$, it can also be shown that using (\ref{anyonandfermion}), the fermionic Hamiltonian (\ref{Hxyincs}) can be transformed to hard-core boson Hamiltonian (\ref{hardcorebosonh}) which itself is the quantum XY spin Hamiltonian. In this way, we obtain the 2d Jordan-Wigner transformation.

\bibliography{refJFU}

\end{document}